\documentclass[11pt,psfig,epsf,bbox]{article} 

\usepackage{amsmath,amssymb,amsthm}
\DeclareMathAlphabet{\mathpzc}{OT1}{pzc}{m}{it}

\begin{document}

\newcommand{\vAi}{{\cal A}_{i_1\cdots i_n}} \newcommand{\vAim}{{\cal
A}_{i_1\cdots i_{n-1}}} \newcommand{\vAbi}{\bar{\cal A}^{i_1\cdots i_n}}
\newcommand{\vAbim}{\bar{\cal A}^{i_1\cdots i_{n-1}}}
\newcommand{\htS}{\hat{S}} \newcommand{\htR}{\hat{R}}
\newcommand{\htB}{\hat{B}} \newcommand{\htD}{\hat{D}}
\newcommand{\htV}{\hat{V}} \newcommand{\cT}{{\cal T}} \newcommand{\cM}{{\cal
M}} \newcommand{\cMs}{{\cal M}^*}
 \newcommand{\vk}{{\bf k}}
\newcommand{\vK}{{\vec K}} \newcommand{\vb}{{\vec b}} \newcommand{{\vp}}{{\vec
p}} \newcommand{{\vq}}{{\vec q}} \newcommand{\vQ}{{\vec Q}}
\newcommand{\vx}{{\vec x}}
\newcommand{\tr}{{{\rm Tr}}} 
\newcommand{\beq}{\begin{equation}}
\newcommand{\eeq}[1]{\label{#1} \end{equation}} 
\newcommand{\half}{{\textstyle
\frac{1}{2}}} \newcommand{\gton}{\stackrel{>}{\sim}}
\newcommand{\lton}{\mathrel{\lower.9ex \hbox{$\stackrel{\displaystyle
<}{\sim}$}}} \newcommand{\ee}{\end{equation}}
\newcommand{\ben}{\begin{enumerate}} \newcommand{\een}{\end{enumerate}}
\newcommand{\bit}{\begin{itemize}} \newcommand{\eit}{\end{itemize}}
\newcommand{\bc}{\begin{center}} \newcommand{\ec}{\end{center}}
\newcommand{\bea}{\begin{eqnarray}} \newcommand{\eea}{\end{eqnarray}}
\newcommand{\beqar}{\begin{eqnarray}} \newcommand{\eeqar}[1]{\label{#1}
\end{eqnarray}} \newcommand{\bra}[1]{\langle {#1}|}
\newcommand{\ket}[1]{|{#1}\rangle}
\newcommand{\norm}[2]{\langle{#1}|{#2}\rangle}
\newcommand{\brac}[3]{\langle{#1}|{#2}|{#3}\rangle} \newcommand{\hilb}{{\cal
H}} \newcommand{\pleft}{\stackrel{\leftarrow}{\partial}}
\newcommand{\pright}{\stackrel{\rightarrow}{\partial}}

\begin{center}
{\Large {\bf{Transition Radiation in QCD matter}}}

\vspace{1cm}

{ Magdalena Djordjevic$^{1,2}$}

\vspace{.8cm}

$^1${\em Department of Physics, Columbia University, New York, NY 10027, USA}

$^2${\em Department of Physics, The Ohio State University, Columbus, OH 43210, 
USA}

\vspace{.5cm}

\today
\end{center}

\vspace{.5cm}

\begin{abstract}
In ultrarelativistic heavy ion collisions a finite size QCD medium is created. 
In this paper we compute radiative energy loss to zeroth order in opacity by 
taking into account finite size effects. Transition radiation occurs on the
boundary between the finite size medium and the vacuum, and we show that it 
lowers the difference between medium and vacuum zeroth order radiative 
energy loss relative to the infinite size medium case. Further, in all 
previous computations of light parton radiation to zeroth order in opacity, 
there was a divergence caused by the fact that the energy loss is infinite in 
the vacuum and finite in the QCD medium. We show that this infinite 
discontinuity is naturally regulated by including the transition radiation.
\end{abstract}

\section{Introduction}

The suppression pattern of high transverse momentum hadrons is considered to 
be a powerful tool to map out the density of the produced QCD 
plasma~\cite{Gyulassy_2002}-\cite{Gyulassy:1991xb}. This suppression 
(so called jet quenching) is assumed to be mainly due to the medium induced 
radiative energy loss of high energy partons propagating through ultra-dense 
QCD matter~\cite{MVWZ:2004}-\cite{KW:2004}. However, even if final state 
multiple elastic and inelastic interactions are neglected, the difference 
between medium and the vacuum $0^{th}$ order energy loss would still be 
significant. This is due to the fact that the gluon dispersion relation is 
different in the medium and the vacuum, leading to differences in
the associated $0^{th}$ order radiation. This effect was first pointed
out by Ter-Mikayelian~\cite{TM1,TM2}, who considered the QED plasma case. 
In~\cite{DG_TM,DG_PLB} we developed a non-abelian QCD analog of the 
Ter-Mikayelian plasmon effect for the case of heavy quarks. We showed that 
while the Ter-Mikayelian effect is negligible for bottom quarks, it has
an important effect on charm quarks, since it leads to a significant reduction 
of the vacuum radiation. This result is a consequence of the fact that the
gluons in the QCD medium acquire a finite mass proportional to the temperature 
of the medium. 

The computation presented in~\cite{DG_TM} was done under the assumption of 
an infinite QCD medium. This raises the question how to generalize these 
results to the more realistic case of a finite size QCD medium created in 
ultra-relativistic heavy ion collisions (URHIC). How the results 
from~\cite{DG_TM} will be modified by finite size QCD effects is the first 
goal of this paper. 

Further, it is well known that light parton $0^{th}$ order energy loss is not 
infrared safe, i.e. it goes to infinity when the parton mass goes to zero. 
This infrared divergence is absorbed in the DGLAP evolution~\cite{Field}, so 
that the only part which contributes to the jet quenching is the difference 
between medium and vacuum energy loss. Since all previous 
computations~\cite{MVWZ:2004,KW:2004} assumed that the light quarks and gluons 
have the same zero mass in both the medium and the vacuum, the difference 
between medium and vacuum energy loss was found to be finite.

However, with the introduction of the Ter-Mikayelian effect, the finite parton 
mass in the medium regulates the infrared divergence of the $0^{th}$ order
energy loss in the medium, while the corresponding energy loss in the vacuum 
remains infinite. This leads to the question how to regulate this 
discontinuity between medium and vacuum light parton energy losses. The 
second goal of this paper is to show how transition radiation naturally 
solves this problem.

The outline of the paper is as follows. In Section~\ref{TR2}, we will compute 
the $0^{th}$ order radiation in a finite size QCD medium for both light and 
heavy quarks. For charm quarks we will show that the transition radiation 
lowers the Ter-Mikayelian effect from $30\%$~\cite{DG_TM} to $15-20 \%$. 
Additionally, we will show that for light partons the transition radiation 
naturally regulates the infinite discontinuity between $0^{th}$ order medium 
and the vacuum radiative energy loss. In Section~\ref{TR3}, we will study how 
the difference between medium and the vacuum $0^{th}$ order energy loss depends
on where the particle is produced. We will show that the difference between 
medium and vacuum $0^{th}$ order energy loss is positive (as intuitively 
expected) as long as the probe is produced far outside the medium (QED
case). However, if the particle is produced inside the medium, such as in the
QCD case, we will obtain that this naive expectation may not hold. In
Section~\ref{TR4}, we will extend our study from Sections~\ref{TR2} 
and~\ref{TR3} to include the fact that due to the confinement in the vacuum, 
the gluons may acquire finite mass. We will obtain qualitatively different 
results, compared to those presented in Sections~\ref{TR2} and~\ref{TR3}, if 
the gluon mass in the vacuum is larger than in the medium. In 
Section~\ref{TR5}, we will combine the results presented in Sections~\ref{TR2} 
and~\ref{TR4} with the medium induced radiative energy 
loss~\cite{Djordjevic:2003zk}.  We find that for certain realistic values of 
the gluon mass in the vacuum, the light quarks can leave the $L<3$~fm medium 
essentially unquenched. We argue that the results presented in the 
Section~\ref{TR5} may provide us with a hint toward solving the puzzle posed 
by~\cite{Mioduszewski}. Finally, in Section~\ref{TR6} we summarize our results 
and put our work in the context of future research.

\section{The one gluon $0^{th}$ order radiation in a finite size QCD medium}
\label{TR2}

The aim of this section is to compute the $0^{th}$ order radiative energy 
loss when the parton is produced in a finite size dielectric medium. To 
introduce the finite size medium, we start from the approach described 
in~\cite{Zakharov}. As in~\cite{Zakharov}, we consider the static medium of 
size $L$, and define two gluon masses, $m_{g,v}$ (for gluon radiated in the 
vacuum), and $m_{g,p}$ (for gluon radiated in the medium). We also assume that,
in general, the running coupling constant can be different in the vacuum and 
in QGP. However, contrary to~\cite{Zakharov}, we ignore spin effects, since 
they are irrelevant in the soft radiation limit that we consider in this paper.

To compute the $0^{th}$ order radiative energy loss in a finite QCD
medium, we have to compute the squared amplitude of a Feynman diagram, 
$M^{rad}$. The Feynman diagram  represents the source $J$, which at time 
${ \mbox x_0}$ produces an off-shell jet with momentum $p^{\prime }$ 
and subsequently (at ${ \mbox x_1}>{ \mbox x_0}$) radiates a gluon 
with momentum $k$. The jet emerges with momentum $p$ and mass $M$. 
We neglect the thermal shifts of the quark mass since 1) for heavy 
quarks, thermal effects on the quark mass are negligible, and 2) 
light quarks will be treated as massless particles for the reason 
explained in footnote 2.   

The matrix element for this $0^{th}$ order in opacity radiation
process can then be written in the following form 
\beqar
M^{rad} =  \int d^4{ \mbox x_0} \; J({ \mbox x_0}) \; d^4 { \mbox x_1}
\; \Delta_M ({ \mbox x_1}-{ \mbox x_0}) \; v^{\mu}({ \mbox x_1}) \; 
A_{\mu}^{\dagger}({ \mbox x_1}) \Phi^{\dagger}({ \mbox x_1})
\eeqar{eq:1}
where $\Phi({ \mbox x_1})=e^{-ip{ \mbox {\scriptsize x}}}$ is the wave
function of the final quark with (on-shell) momentum $p$ and 
$A_\mu({ \mbox x_1})$ is the wave function of the emitted gluon. 
Vertex function $v^{\mu}({ \mbox x_1})$ is given by $v^{\mu}({ \mbox x_1})
=g({ \mbox x_1})(\overleftarrow{\partial}^{\mu}-
\overrightarrow{\partial}^{\mu})$, where $g({ \mbox x_1})$ is 
the running coupling constant which is in general different in the 
medium than in the vacuum.

For this problem it is convenient to use light cone 
coordinates~\cite{Kogut_1970}. This coordinate system is appropriate for 
systems moving with almost the speed of light. It is obtained by choosing new 
space-time coordinates $[{ \mbox x^{+}}, { \mbox x^{-}}, {\bf
    x}]$,\footnote{Note that the ${ \mbox x^{+}}$ and ${ \mbox x^{-}}$ axes of 
the new frame lie on the light cone~\cite{Kogut_1970}.} related to the 
coordinates in the laboratory frame $(t, z, {\bf x})$ by (${\bf x}$ is
the transverse coordinate)
\beqar
{ \mbox x}^{+}=(t+z) \, , \hspace*{2cm}  { \mbox x}^{-}=(t-z).
\eeqar{xpxm}
In the same way the light cone momentum $[p^{+}, p^{-}, {\bf p}]$ is
related to the momentum in the laboratory frame 
$[E, p_z, {\bf p}]$ by (${\bf p}$ is the transverse momentum)

\beqar
p^{+}=(E+p_z) \, , \hspace*{2cm}  p^{-}=(E-p_z).
\eeqar{pppm}
Additionally, it can be shown that in the light cone coordinate system the 
propagator $\Delta_M({ \mbox x})$ reduces to (see~\cite{Kogut_1970}):

\beqar
\Delta_M({ \mbox x})=\frac{-i}{(2 \pi)^3} \int \frac{dp'^+ d^2 {\bf
    p'}}{2 p'^+} (\theta({ \mbox x}^+)e^{-ip'{ \mbox {\scriptsize x}}}+
\theta(-{ \mbox x}^+)e^{ip'{ \mbox {\scriptsize x}}}),
\eeqar{propagator}
where $p'^{-} \equiv \frac{{\bf p'}^2+M^2}{p'^+}$.

\bigskip
 
In the spinless case, the wave function of the emitted gluon with
momentum $k$ can be written as
\beqar
A_\mu(x) = \epsilon_\mu (k)\;  \Phi_g ({ \mbox x})\; c,
\eeqar{gluon1}
where $\epsilon(k)=[0,2 \frac{{\mbox{\scriptsize{\boldmath $\epsilon$}}} 
{\bf \cdot k}}{k^+}, {\mbox{\scriptsize{\boldmath $\epsilon$}}}]$ is the 
transverse polarization and $c$ is the color factor of the radiated 
gluon. 
\beqar
\Phi_g ({ \mbox x})=e^{-i\frac{1}{2} [k^{+}{ \mbox {\scriptsize x}^{-}}+ 
\int\limits_{0}^{ \mbox {\scriptsize x}^{+}} d\xi \, 
k^{-}(\xi)] + i {\bf k \cdot x}}
\eeqar{Phi}
is the wave function (derived in Appendix A) that satisfies the 
Klein-Gordon equation with position dependent gluon mass $m_g({ \mbox x^+})$,
and $k^{-}({ \mbox x^+}) = \frac{{\bf k}^2+m_g^2
({ \mbox {\scriptsize x}^{+}} )}{k^+}$. In the static approximation 
$m_g({ \mbox x^+})$ becomes
\beq
m_g({ \mbox x^+})=m_{g,p} \; \theta(L-\frac{{ \mbox x^+}}{2})+ 
m_{g,v} \; \theta(\frac{{ \mbox x^+}}{2}-L)\; ,
\eeq{mg}
where $m_{g,p}$ is gluon mass for the gluon radiated in the medium, 
while $m_{g,v}$ is gluon mass for the gluon radiated in the vacuum.

We can now compute $M_{rad}$ by substituting 
Eqs.~(\ref{propagator})-(\ref{mg}) in Eq.(\ref{eq:1}), which leads to the 
fallowing result (see Appendix B for detailed calculation of $M_{rad}$):  
\beqar
M_{rad}=-2 i\, J(p+k)\frac{{\mbox{\boldmath $\epsilon$}}{\bf \cdot k}}{x} 
\left [\frac{g_p}{p'^+}\frac{1-e^{i\chi_p L}}{\chi_p} +
\frac{g_v}{p'^+}\frac{e^{i\chi_p L}}{\chi_v} \right] c \; ,
\eeqar{MradStat}
where $g_p$ ($g_v$) is the running coupling constant in the medium
(vacuum). The variable $x $ is defined as $x \equiv \frac{k^+}{p'^+}$ and 
\beqar
\chi_v&=&\frac{{\bf k}^2+M^2 x^2 + m_{g,v}^2}{xp'^+} \; , \nonumber \\
\chi_p&=& \frac{{\bf k}^2+M^2 x^2 + m_{g,p}^2}{xp'^+} \; .
\eeqar{eq:chi}
Here, we use for the initial quark a plane wave state in the $ {\bf x}$-plane 
and set ${\bf p'}=0$. Then ${\bf p}=- {\bf k}$, $p'^-=\frac{M^2}{p'^+}$ and 
$p^-=\frac{{\bf k}^2+M^2}{(1-x)p'^+}$. Note that soft radiation is defined as 
$x\ll 1$ (i.e. $p^+\gg k^+$), so we assume that 
$1-x \approx 1$.\footnote{Note that in the light quark case we set $M=0$~GeV. 
Keeping the finite light quark mass in Eq.~(\ref{eq:chi}) would correspond 
to retaining small $x^2$ corrections, which would be inconsistent with 
$1-x \approx 1$.} Additionally, as in~\cite{GLV}, we assume that $J$
varies slowly with $p$, so that $J(p+k) \approx J(p)$. 

\medskip

In soft gluon approximation, the spectrum can be extracted from 
Eq.~(\ref{MradStat}) as (see~\cite{GLV})
\beqar
|M_{rad}^{0}|^{2} \frac{d^{3} \vec{{\bf p}}}{2 E (2 \pi)^{3}}
\frac{d^{3} \vec{{\bf k}}}{2 \omega (2 \pi)^{3}} \approx 
d^{3}N_J \; d^{3} N_g^{(0)}
\; \; ,
\eeqar{imm1a} 
where 
\beqar
d^{3} N_{J} = d_{R} |J(p)|^{2} 
\frac{d^{3}\vec{\mathbf{p}}}{( 2\pi )^{3}2p^{0}} \; \; .
\eeqar{FO2}
Here $d_R=3$ (for three dimensional representation of the quarks).

Finally, by using Eqs.~(\ref{imm1a},~\ref{FO2}) together with 
Eq.~(\ref{MradStat}) we obtain the main order fractional energy loss 
$(I\equiv \Delta E/E)$ for massive quarks and gluons in the QCD medium 
of finite size L
\beqar
\frac{dI_{med}^{(0)}}{dx \; d^2{\bf k}}  & =& \frac{dI_{vac}^{(0)}}
{dx \; d^2{\bf k}} +2 \frac{ C_R {\bf k}^2 }{\pi^2 (p'^+)^2}  
\frac{ \sqrt{\alpha_{S}^p(k)}}{\chi_p}
\left( \frac{ \sqrt{\alpha_{S}^p(k)}}{\chi_p} - 
\frac{\sqrt{\alpha_{S}^v(k)}}{\chi_v} \right) (1-\cos (\chi_p L))
\nonumber \\ &\xrightarrow{ \alpha_{S}^p=\alpha_{S}^v} &
\frac{dI_{vac}^{(0)}}
{dx \; d^2{\bf k}} +2 \frac{ C_R \alpha_{S} }{\pi^2}  
\frac{ {\bf k}^2 (m_{g,v}^2- m_{g,p}^2) }{({\bf k}^2+m_{g,p}^2 +M^2x^2)^2
({\bf k}^2+m_{g,v}^2 +M^2x^2)}\nonumber \\ && \hspace*{3cm}
\times(1-\cos (\frac{({\mbox{{\bf k}}}^2+m_{g,p}^2 +M^2x^2)L}{2E x})),
\eeqar{eq:9}
where $\alpha_{S}^p(k)$ ($\alpha_{S}^v(k)$) is the running coupling constant 
in the medium (vacuum). $C_R$ is the color Casimir for the partons, i.e. 
$C_R=4/3$ for quarks and $C_R=3$ for gluons. $E$ is the initial jet
energy. The second equation in~(\ref{eq:9}) is valid in the 
$\alpha_{S}^p(k)=\alpha_{S}^v(k)$ case, where the physical meaning of the 
obtained results are more evident. 

$I_v^{(0)}$ is $0^{th}$ order in opacity fractional energy loss in the vacuum:
\beqar
\frac{d I_{vac}^{(0)}}{dx \; d^2{\bf k}}= \frac{C_R \alpha_{S}^v(k)}{\pi^2} 
\frac{{\bf k}^2}{({\bf k}^2+ m_{g,v}^2 +M^2x^2)^2} \; 
\eeqar{eq:vac}
and $I_{med}^{(0)}$ is total $0^{th}$ order in opacity fractional energy loss 
in finite size medium, given by   
\beqar
I_{med}^{(0)}= I_{TM}^{(0)} +I_{trans}^{(0)} \; .
\eeqar{eq:10b}
Here, $I_{TM}^{(0)}$ is the $0^{th}$ order in opacity fractional energy loss 
in the infinite size medium, which can be obtained from $I_{vac}^{(0)}$ by 
replacing $m_{g,v}$ by $m_{g,p}$ in Eq.~(\ref{eq:vac}). $I_{trans}^{(0)}$ 
is the additional {\em transition radiation} occurring when the jet is 
traversing from the medium to the vacuum. As a crosscheck, we note that, by 
neglecting spin effects in~\cite{Zakharov}, Eq.~(9) from~\cite{Zakharov} 
can be reduced to the Eq.~(\ref{eq:9}) above. We note that the computation
in~\cite{Zakharov} was done in 3-dimensional coordinate space ($z$, {\bf x}),
while our computations were more consistently performed in the light-cone 
4-dimensional coordinate space.    

To obtain the fractional energy losses, we perform the integration by 
using $0<|{\bf k}|< 2 x (1-x) E$, i.e. in our computations we 
do not introduce a lower momentum cutoff. For running coupling we use the 
"Frozen $\alpha$ model"~\cite{Dokshitzer-FrozenAlpha} 
\beq
\alpha_{S}(Q^{2}-M^{2}) = {\rm Min} \{ 0.5, \frac{4 \pi}
{\beta_{0} Log(\frac{Q^{2}-M^{2}}{\Lambda_{QCD}^{2}})} \} =
{\rm Min} \{ 0.5, \frac{4 \pi} {\beta_{0} 
Log(\frac{{\bf k}^2+ m_{g}^2 +M^2x^2}{x \Lambda_{QCD}^{2}})} \}\ \;\;\;,
\eeq{frozen_alpha}
where $\beta_{0}=\frac{28}{3}$ for effective number of flavors 
$n_{f} \approx 2.5$, $\Lambda_{QCD} \approx 0.2$~GeV, and 
$Q^{2}-M^{2} = \frac{{\bf k}^2+ m_g^2 +M^2x^2}{x}$. Note that
$\alpha_{S}^p$ ($\alpha_{S}^v$) is obtained by setting
$m_{g}^2=m_{g,p}^2$ ($m_{g}^2=m_{g,v}^2$) in Eq.~(\ref{frozen_alpha}).  
\medskip

\begin{figure}
\vspace*{6.9cm} \includegraphics{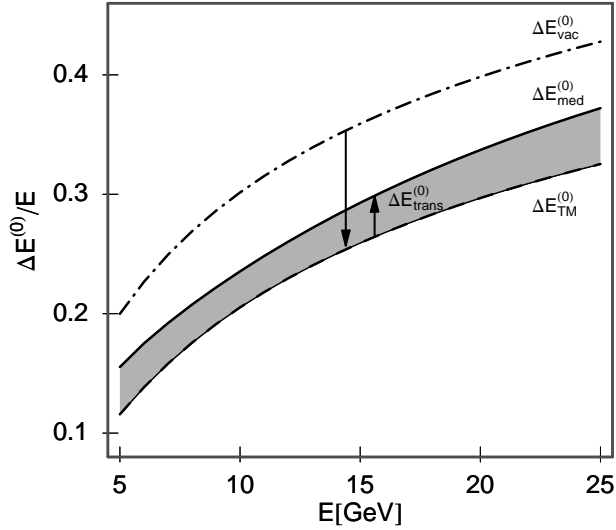}
\caption{ The reduction of the fractional $0^{th}$ order energy loss for 
charm quarks due to both the QCD Ter-Mikayelian effect and transition 
radiation is shown as a function of the quark energy. The dashed-dotted 
curve shows the vacuum energy loss if gluons are treated as massless and
transversely polarized. Dashed curve shows the effect using the Ter-Mikayelian 
effect only, and solid curve shows the total effect by including both
the Ter-Mikayelian effect and transition radiation. Assumed thickness of the 
medium is $L=5$~fm, charm mass is $M=1.5$~GeV and the gluon mass is 
$m_{g,v}=0$~GeV ($m_{g,p} =0.35$~GeV) for the vacuum (medium) case. }
\label{fig:TR1}
\end{figure}

\begin{figure}
\vspace*{6.9cm} \includegraphics{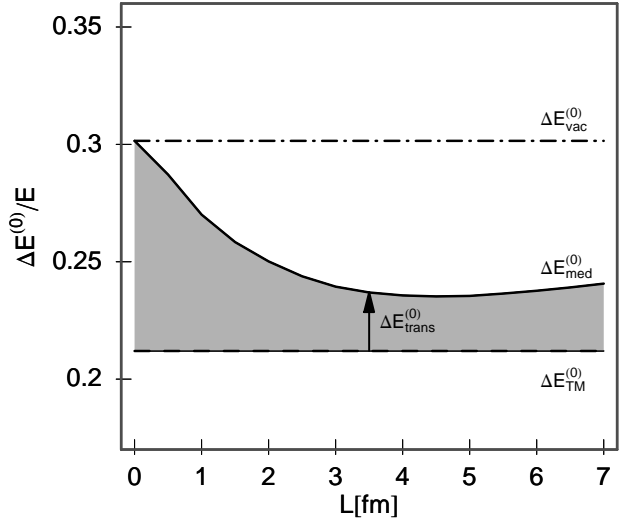}
\caption{ The reduction of the fractional $0^{th}$ order energy loss for 
$10$~GeV charm jet due to both the QCD Ter-Mikayelian effect and transition 
radiation is shown as a function of the thickness of the medium. The 
dashed-dotted curve shows the vacuum energy loss if gluons are treated as 
massless and transversely polarized. Dashed curve shows the effect using the 
Ter-Mikayelian effect only, and solid curve shows the total effect by 
including both the Ter-Mikayelian effect and transition radiation. We assume 
that the mass of the charm jet is $M=1.5$~GeV and the gluon mass is 
$m_{g,v}=0$~GeV ($m_{g,p} =0.35$~GeV) for the vacuum (medium) case.}
\label{fig:TR2}
\end{figure}

We next use Eqs.~(\ref{eq:9})-(\ref{frozen_alpha}) to provide a
comprehensive analysis of the influence of the transition radiation to
the total $0^{th}$ order in opacity energy loss for both light and
heavy quarks. We first concentrate on the charm quark case and look how 
the $0^{th}$ order energy loss depend on the initial jet energy  
(Fig.~\ref{fig:TR1}) and thickness of the medium (Fig.~\ref{fig:TR2}). In 
Fig.~\ref{fig:TR1} we see that for charm quarks, transition radiation lowers 
the Ter-Mikayelian effect from $30\%$ to $15-20 \%$ for $L=5$~fm medium. In 
Fig.~\ref{fig:TR2} we see that for a medium thickness greater than 4~fm 
the transition radiation becomes approximately independent of the thickness 
of the medium.

\begin{figure}
\vspace*{6.4cm} \includegraphics{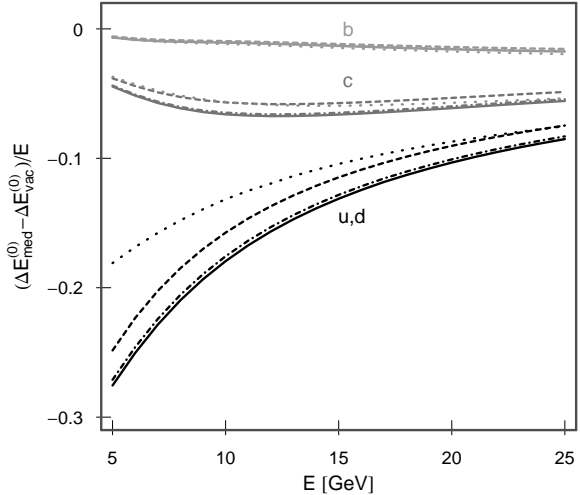}
\caption{ The difference between medium and the vacuum $0^{th}$ order 
fractional energy loss for light (the lower four curves), charm (the middle 
four curves) and bottom quarks (the upper four curves) is shown as a function 
of initial jet energy. Solid curves are computed by assuming running coupling 
given by Eq.~(\ref{frozen_alpha}), with {\em different} $\alpha_{S}^p$ in the 
medium and $\alpha_{S}^v$ in the vacuum. Dashed, dot-dashed and dotted curves 
are computed by assuming the same coupling constant both in the medium and in 
the vacuum. For dashed (dot-dashed) curves we used running coupling 
$\alpha_{S}^p$ ($\alpha_{S}^v$) given by Eq.~(\ref{frozen_alpha}). Dotted 
curves are computed by assuming constant coupling $\alpha_{S}=0.3$. Assumed 
thickness of the medium is $L=5$~fm, light quark mass is $M=0$~GeV, charm mass 
$M=1.5$~GeV, and bottom mass $M=4.5$~GeV. The gluon mass is $m_{g,v}=0$~GeV 
($m_{g,p}=0.35$~GeV) for the vacuum (medium) case.}
\label{fig:TR_LCB}
\end{figure}

The previous two figures were computed by assuming running coupling (given by 
Eq.~(\ref{frozen_alpha})) which is different in the medium and in the vacuum. 
In Fig.~\ref{fig:TR_LCB} we want to test how the obtained medium and the vacuum
fractional energy loss difference ($I_{med}^{(0)}- I_{vac}^{(0)}$) is robust 
against variations in the choice of coupling constant. To do that, we plot 
$I_{med}^{(0)}- I_{vac}^{(0)}$ as a function of initial jet energy for
three different choices of running coupling as well as constant coupling 
$\alpha_{S}=0.3$. We see that for heavy ($c$ and $b$) quarks, the difference 
between medium and the vacuum energy loss is almost independent on the choice 
of coupling constant. For the light quark case, we see that while 
$I_{med}^{(0)}- I_{vac}^{(0)}$ is robust to the choice of running coupling, 
it is fairly sensitive to the choice between running and constant coupling. 
For example, in $p_\perp<10$~GeV range, $30\%$ smaller 
$I_{med}^{(0)}-I_{vac}^{(0)}$ is obtained when constant coupling 
$\alpha_{S}=0.3$ is employed.

Additionally, from Fig.~\ref{fig:TR_LCB} we see that the finite mass (a.k.a.
{\em dead cone}~\cite{Dokshitzer:2001zm}) effect on 
$I_{med}^{(0)}-I_{vac}^{(0)}$ is strong, i.e. we see a qualitative difference 
between light and heavy quark $0^{th}$ energy losses, which persists at high 
momentum. For example, while $I_{med}^{(0)}-I_{vac}^{(0)}$ for light
quarks is large and shows a noticeable dependence on jet energy, it is 
negligible for bottom quarks in the whole jet energy range.

The most striking observation from Fig.~\ref{fig:TR_LCB} is that, for the 
light quark case, the difference between medium and the vacuum energy loss 
($I_{med}^{(0)}-I_{vac}^{(0)}$) is finite. To validate this numerical result 
analytically, we assume a perturbative vacuum (i.e. $m_{g,v}=0$~GeV) and 
the same coupling in the medium and in the vacuum (i.e. 
$\alpha_{S}^p=\alpha_{S}^v$). In the ($M=0$~GeV) light quark case, the 
Eq.~(\ref{eq:9}) reduces to

\beqar
\frac{d(I_{med}^{(0)}-I_{vac}^{(0)})}{dx \; d^2{\bf k}}  = 
- 2 \frac{ C_R \alpha_{S} }{\pi^2}  
\frac{ m_{g,p}^2 }{({\bf k}^2+m_{g,p}^2)^2}
\left(1-\cos (\frac{({\mbox{{\bf k}}}^2+m_{g,p}^2)L}{2E x}) \right),
\eeqar{eq:Light}
This equation is infrared safe when ${\bf k}\rightarrow 0$. This is an 
important result having in mind that without transition radiation, the 
Ter-Mikayelian effect leads to a discontinuity between finite medium
and infinite vacuum energy loss. Therefore, we conclude that the transition 
radiation has special importance in the case of the light quarks since it 
provides a natural regularization of medium dispersion effects.

\begin{figure}
\vspace*{6.9cm} \includegraphics{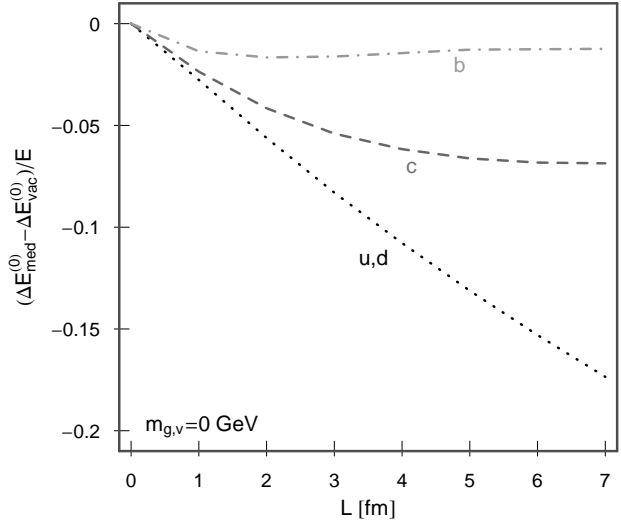}
\caption{ The difference between medium and the vacuum $0^{th}$ order
fractional energy loss for light (dotted curve), charm (dashed curve) 
and bottom quark (dot-dashed curve) is shown as a function of the
thickness of the medium. The initial jet energy is $E=15$~GeV. The
curves were computed by assuming running coupling given by 
Eq.~(\ref{frozen_alpha}), with different coupling in the medium and 
in the vacuum. Light quark mass is $M=0$~GeV, charm mass $M=1.5$~GeV, 
and bottom mass $M=4.5$~GeV. The gluon mass is $m_{g,v}=0$~GeV ($m_{g,p}
=0.35$~GeV) for the vacuum (medium) case.}
\label{fig:TR_Ldep}
\end{figure}

In Fig.~\ref{fig:TR_Ldep} we fix the energy jet to $15$~GeV, and test
how the $I_{med}^{(0)}- I_{vac}^{(0)}$ depends on $L$ for different
types of quarks. We see that light and heavy quarks show significantly
different thickness dependence. For heavy quarks, 
$I_{med}^{(0)}- I_{vac}^{(0)}$ saturate after some value of $L$
(i.e. after $L=4$~fm for charm and $L=1$~fm for bottom). This saturation 
behavior is expected having in mind that for massive quarks and large
enough $L$, $\cos(\chi_p L)$ becomes rapidly oscillating function. Then 
$\left\langle \cos(\chi_p L) \right\rangle \rightarrow 0$, and 
Eq.~(\ref{eq:9}) becomes independent on the thickness of the medium. For the 
light quarks, we see that $|I_{med}^{(0)}- I_{vac}^{(0)}|$ increases
approximately linearly with $L$. This linear thickness dependence persists for 
higher jet energies as well (results now shown). This result is unexpected,
having in mind that, in the light quark case and $L/E \rightarrow 0$ limit, by 
expanding cosine in Eq.~(\ref{eq:Light}), we expect to obtain quadratic ($L^2$)
thickness dependence for $I_{med}^{(0)}- I_{vac}^{(0)}$. However, by 
integrating Eq.~(\ref{eq:Light}) in the $L/E \rightarrow 0$ limit (and without 
expanding cosine) we obtain a result which depends linearly on the 
thickness $L$ of the medium (in agreement with the numerical results shown in 
Fig.~\ref{fig:TR_Ldep}):
\beqar
I_{med}^{(0)}-I_{vac}^{(0)}
\approx
\frac{C_R \alpha_S}{2} \frac{m_{g,p}^2 L}{E} \ln[\frac{E}{2 m_{g,p}}].
\eeqar{L_dep_light}

Finally, from Figs.~\ref{fig:TR1}-\ref{fig:TR_Ldep} we see that the total
energy loss in the medium is smaller than in the vacuum. This result comes 
from Eq.~(\ref{eq:9}), where we see that $\Delta E_p^{(0)}-\Delta E_v^{(0)} 
\propto (m_{g,v}^2-m_{g,p}^2)$. Therefore, if the gluon mass in the medium is 
larger than in the vacuum, then the total $0^{th}$ order energy loss in the 
medium will be smaller than in the vacuum. Though mathematically correct, this 
result seems surprising, since it would be expected that the radiation
in the medium (even at the lowest order) is always larger than the 
corresponding radiation in the vacuum, particularly having in mind the
work presented in~\cite{TM2,Jackson}. In~\cite{TM2,Jackson} the transition 
radiation was studied for the particle traversing the QED medium, and it was 
shown that the lowest order radiative energy loss in the medium is always 
larger than in the vacuum. This study considered the case when the particle is 
produced outside the medium (at $z_{0}=-\infty$). However, contrary to
the usual experiments involving a QED medium, where the medium is probed
by using test particles produced far outside the medium, in URHIC the probes 
are produced {\em inside} the medium. Therefore, to be able to intuitively 
understand the result obtained from Eq.~(\ref{eq:9}), we have to take into 
account the qualitative change in the experimental approaches between QED and 
QCD. In the next Section we study how the difference between medium and the 
vacuum energy loss depends on the point of particle production.

\section{Dependence of the $0^{th}$ order energy loss on the point of 
particle production}
\label{TR3}

\begin{figure}[h]
\vspace*{5.3cm} 
\includegraphics{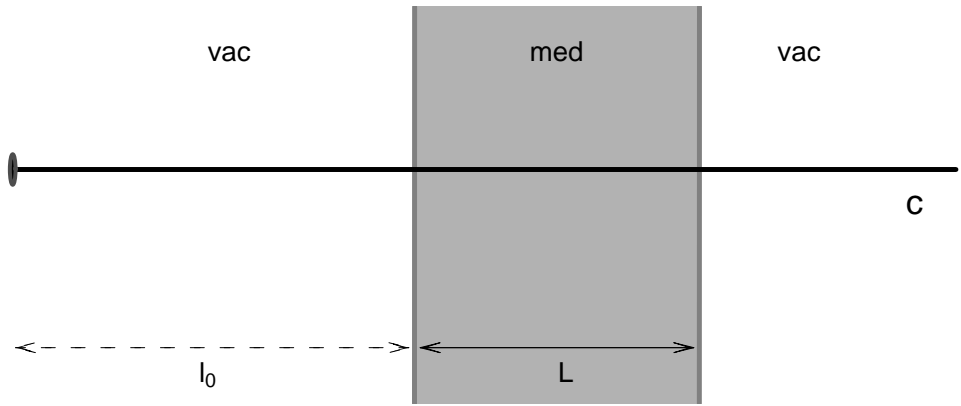}
\caption{An illustration of the system studied in this Section. Gray 
area shows the medium of thickness $L$. Probe is produced in the vacuum 
at distance $|l_0|$ from the medium.}
\label{fig:TR3}
\end{figure}
To study this particular problem we assume the static system shown in
Fig.~\ref{fig:TR3}. That is, we consider the probe produced in the vacuum at 
the finite distance $l_0$ (i.e. ${\mbox x}_0^+=-2l_0<0$) from the medium of 
size $L$.  As in the previous section, we assume that a gluon is subsequently 
radiated at a point ${\mbox x}_1$. A gluon radiated in the vacuum (medium) has 
the mass $m_{g,v}$ ($m_{g,p}$). Therefore, the gluon mass 
$m_g({ \mbox x_1^+})$ has the following form
\beq
m_g({ \mbox x}_1^+)=m_{g,v}\; \theta(-{\mbox x}_1^+) +
m_{g,p} \; \theta({\mbox x}_1^+) \; \theta(L-\frac{{ \mbox x}_1^+}{2})+ 
m_{g,v} \; \theta(\frac{{ \mbox x}_1^+}{2}-L).
\eeq{mg_1}

As in the Section~\ref{TR2}, we can now compute $M_{rad}$, by 
substituting Eqs.~(\ref{propagator})-(\ref{Phi}) and~(\ref{mg_1}) in 
Eq.(\ref{eq:1}), which leads to the fallowing result (for the derivation 
see Appendix C):  

\beqar
M_{rad} = -2 i\, J(p+k)\frac{{\mbox{\boldmath $\epsilon$}}{\bf \cdot k}}{x} 
\frac{1}{p'^+}\left [\frac{g_v}{\chi_v} -(\frac{g_v}{\chi_v}-
\frac{g_p}{\chi_p})e^{i\chi_v l_0}(1-e^{i\chi_p L})\right ].
\eeqar{Mrad_outside}
Eq.~(\ref{Mrad_outside}) together with the Eqs.~(\ref{imm1a}) and~(\ref{FO2}), 
leads to

\beqar
\frac{d(I_{med}^{(0)}-I_{vac}^{(0)})}{dx \; d^2{\bf k}}  & =& 
2 \frac{ C_R \; {\bf k}^2 }{\pi^2 (p'^+)^2} 
\left(\frac{\sqrt{\alpha_{S}^p(k)}}{\chi_p}-
\frac{\sqrt{\alpha_{S}^v(k)}}{\chi_v}\right)^2 
(1- \cos (\chi_p L) ) \nonumber \\ 
&-& 2 \frac{ C_R \; {\bf k}^2 }{\pi^2 (p'^+)^2}
\frac{\sqrt{\alpha_{S}^v(k)}}{\chi_v}
\left(\frac{\sqrt{\alpha_{S}^v(k)}}{\chi_v}-
\frac{\sqrt{\alpha_{S}^p(k)}}{\chi_p}\right)
(\cos (\chi_v l_0) -
\cos (\chi_v l_0 + \chi_p L)) .
\eeqar{eq:I_med_outside}

Equation~(\ref{eq:I_med_outside}) represents the difference between
medium and vacuum fractional energy loss when the probe is produced
outside the medium. It is useful to look at two important limits of this 
equation:
 
1) $l_0 \rightarrow 0$. In this limit we recover the case when the particle 
is produced in the medium of size $L$. In this case, the
Eq.~(\ref{eq:I_med_outside}) reduces to Eq.~(\ref{eq:9}) from the
previous section. 

2) $l_0 \rightarrow \infty$. This limit corresponds to the case when the 
particle is produced far outside the medium, i.e. it is equivalent to the 
QED case studied in~\cite{TM2}.

When $l_0 \rightarrow \infty$ the second term in Eq.~(\ref{eq:I_med_outside})
goes to zero. For $\alpha_{S}^p(k)=\alpha_{S}^v(k)$ we obtain a result which 
agrees with~\cite{TM2}:

\beqar
\frac{d(I_{med}^{(0)}-I_{vac}^{(0)})}{dx \; d^2{\bf k}} 
\arrowvert_{l_0 \rightarrow \infty}  
&\xrightarrow{ \alpha_{S}^p=\alpha_{S}^v} &
 2 \frac{ C_R \alpha_{S} (k)} {\pi^2} 
\frac{{\bf k}^2 (m_{g,p}^2-m_{g,v}^2)^2}
{({\bf k}^2+M^2 x^2 + m_{g,v}^2)^2({\bf k}^2+M^2 x^2 + m_{g,p}^2)^2}
\nonumber \\ && \hspace*{1.5cm}\times 
[1-\cos(\frac{({\bf k}^2+M^2 x^2 + m_{g,p}^2)L}{2E x})]\; 
\eeqar{eq:15}

We see that, when $l_0 \rightarrow \infty$, Eq.~(\ref{eq:I_med_outside}) 
becomes positive definite independently of the gluon masses. 
Therefore, in this case the $0^{th}$ order energy loss in the medium is always 
larger than in the vacuum. This result agrees with our intuitive expectations 
and with~\cite{TM2}. 

\begin{figure}
\vspace*{7.3cm} \includegraphics{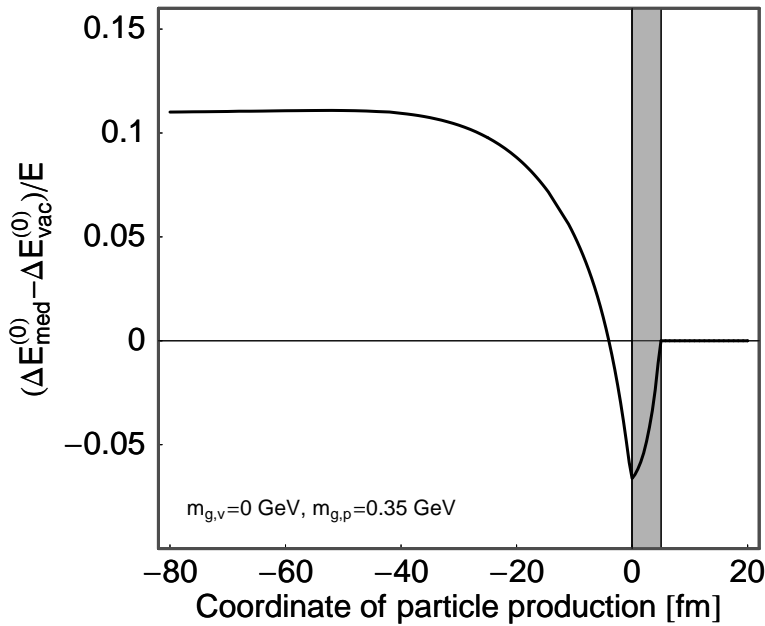}
\caption{ The difference between medium and the vacuum $0^{th}$ order energy 
loss for charm quarks is shown as a function of coordinate of particle 
production. The curves are computed by assuming running coupling given by 
Eq.~(\ref{frozen_alpha}), with different couplings in the medium and 
in the vacuum. The thickness of the medium is $L=5$~fm. The charm mass is 
$M=1.5$~GeV and the gluon mass is $m_{g,v}=0$~GeV ($m_{g,p} =0.35$~GeV)  for 
the vacuum (medium) case. }
\label{fig:TR4}
\end{figure}
\vskip 4truemm 

Finally, in Fig.~\ref{fig:TR4} we show the difference between medium and 
vacuum $0^{th}$ order energy loss as a function of coordinate of particle 
production. The figure clearly shows the transition from positive definite 
values (for the case when the particle is produced far outside the medium) to 
negative values (obtained when the particle is produced inside the medium).

\newpage

\section{Dependence of transition radiation on the vacuum gluon mass}
\label{TR4}

\begin{figure}
\vspace*{6.3cm} \includegraphics{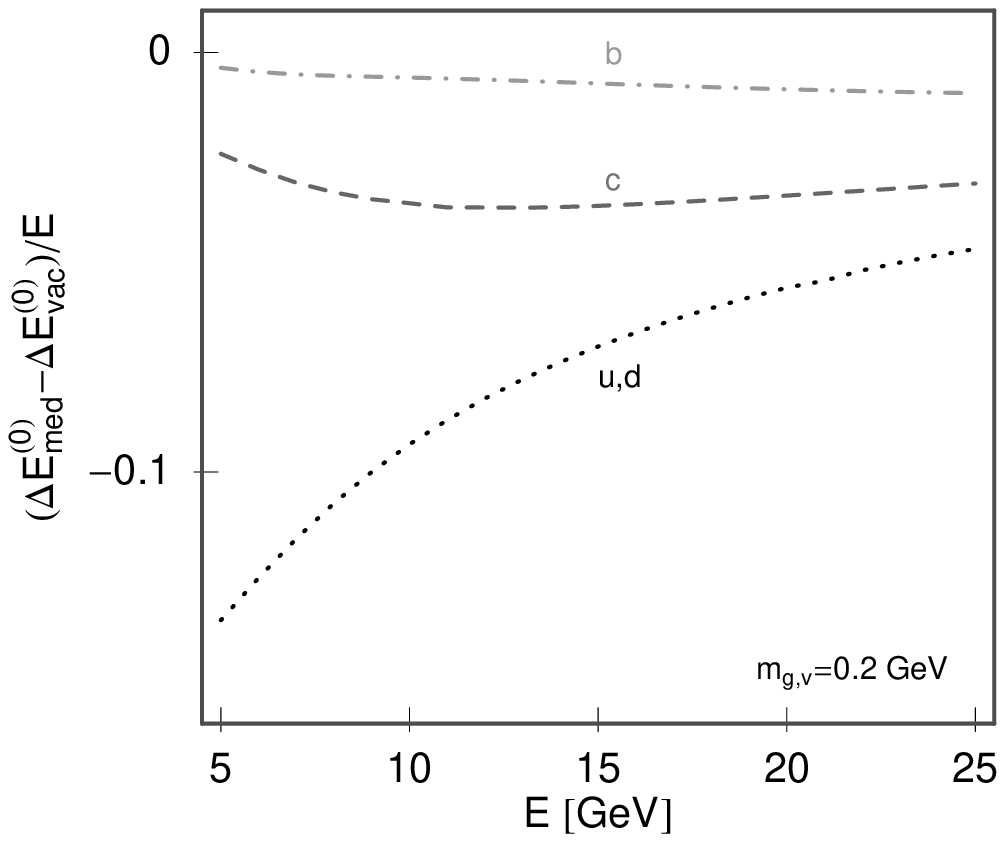}
\includegraphics{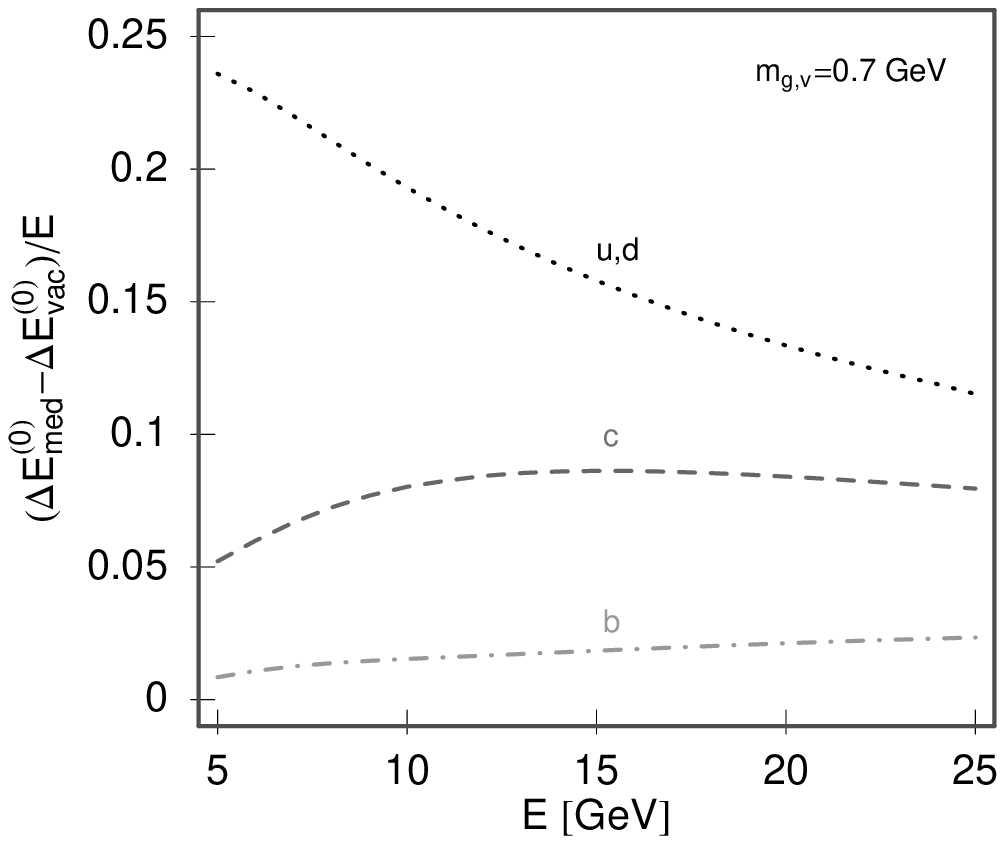}
\caption{ The difference between medium and the vacuum $0^{th}$ order
fractional energy loss for light (dotted curve), charm (dashed curve) 
and bottom quark (dot-dashed curve) is shown as a function of the
jet energy. The thickness of the medium is $L=5$~fm. The
curves are computed by assuming running coupling given by 
Eq.~(\ref{frozen_alpha}), with different couplings in the medium and 
in the vacuum. Light quark mass is $M=0$~GeV, charm mass $M=1.5$~GeV, 
and bottom mass $M=4.5$~GeV. The gluon mass in the medium is 
$m_{g,p} = 0.35$~GeV. Left (right) panel corresponds to the 
$m_{g,v}=\Lambda_{QCD}\approx 0.2$~GeV ($m_{g,v}=0.7$~GeV) case ($m_{g,v}$ 
is the gluon mass in the vacuum).}
\label{fig:TR_LCB_mgv}
\end{figure}
\vskip 4truemm 

The analysis presented in the previous two sections is based on the assumption 
that $m_{g,v}=0$ in URHIC. However, this is true only in the case of the 
perturbative vacuum which does not take confinement into account. A
phenomenological way to simulate confinement in the vacuum is to introduce an 
effective gluon mass $m_{g,v} \neq 0$~\cite{Nikolaev}. In this case, there 
are two different vacuum gluon masses that can be found in 
literature~\cite{Sterman}-\cite{Field_02}, i.e. 
$m_{g,v} \approx \Lambda_{QCD}$ and $m_{g,v}\approx 0.7$~GeV. 

\begin{figure}
\vspace*{6.3cm} \includegraphics{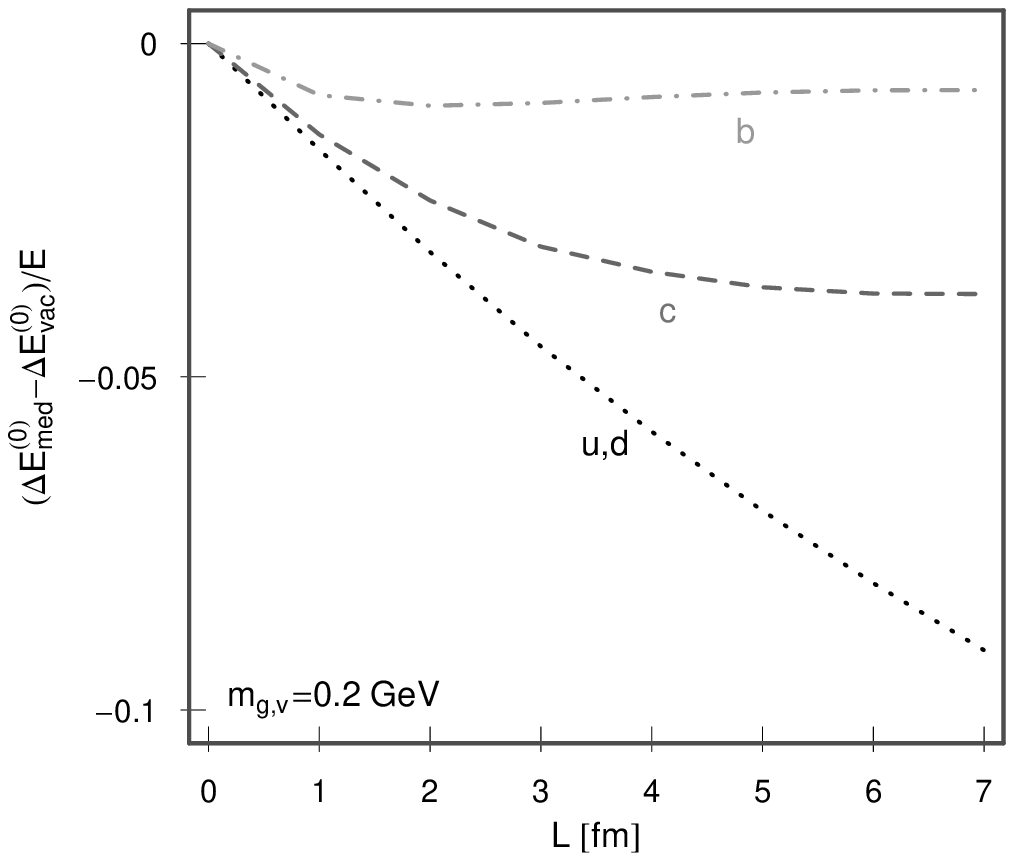}
\includegraphics{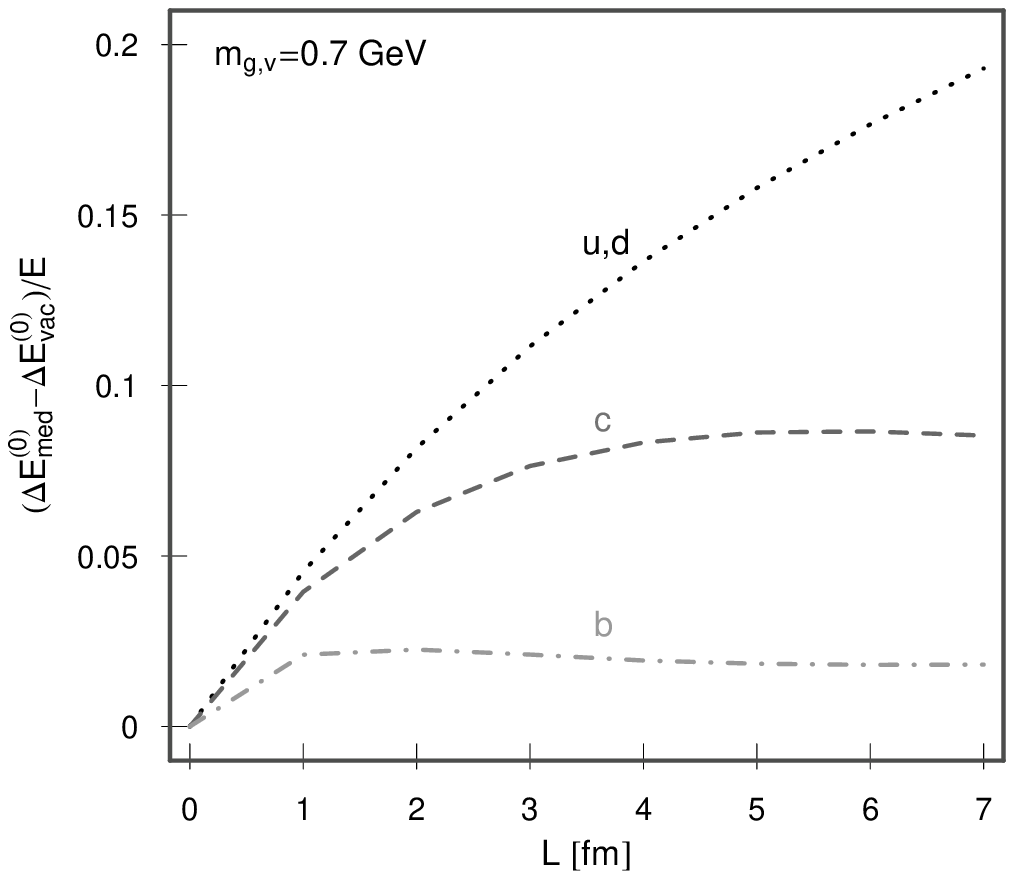}
\caption{ The difference between medium and the vacuum $0^{th}$ order
fractional energy loss for light (dotted curve), charm (dashed curve) 
and bottom quark (dot-dashed curve) is shown as a function of the
thickness of the medium. The initial jet energy is $E=15$~GeV. The
curves are computed by assuming running coupling given by 
Eq.~(\ref{frozen_alpha}), with different couplings in the medium and 
in the vacuum. Light quark mass is $M=0$~GeV, charm mass $M=1.5$~GeV, 
and bottom mass $M=4.5$~GeV. The gluon mass in the medium is 
$m_{g,p} = 0.35$~GeV. Left (right) panel corresponds to the 
$m_{g,v}=\Lambda_{QCD}\approx 0.2$ ($m_{g,v}=0.7$)~GeV case ($m_{g,v}$ 
is the gluon mass in the vacuum).}
\label{fig:TR_Ldep_mgv}
\end{figure}
\vskip 4truemm

\begin{figure}
\vspace*{6.3cm} \includegraphics{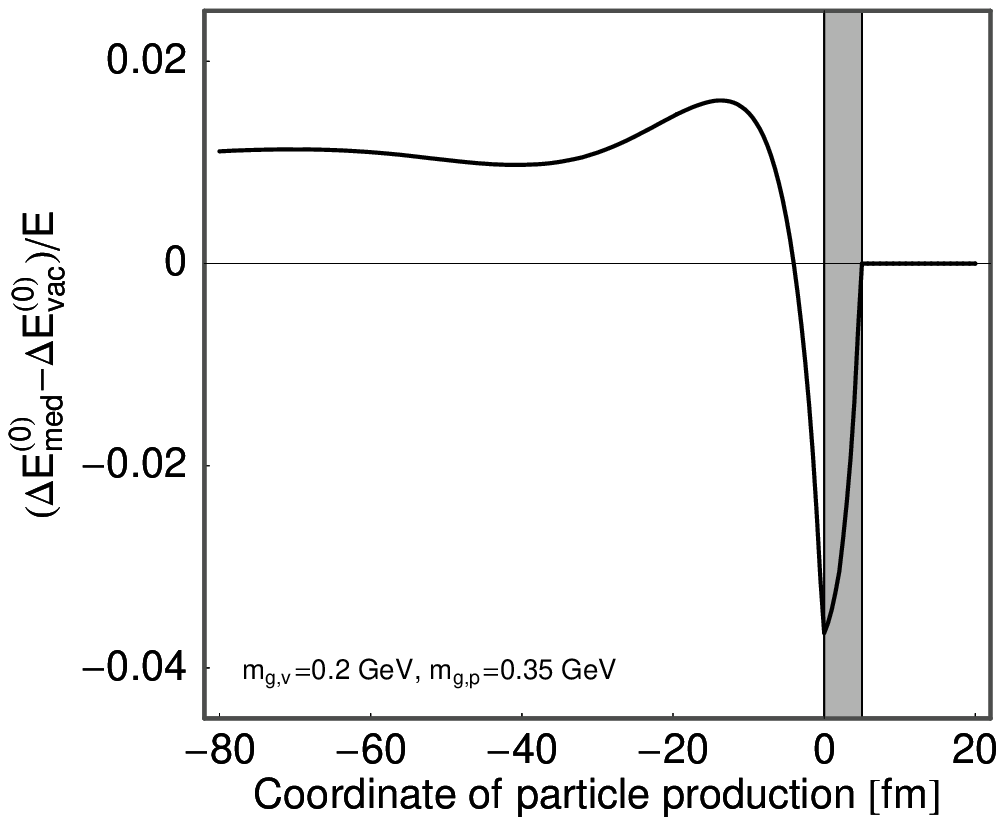}
\includegraphics{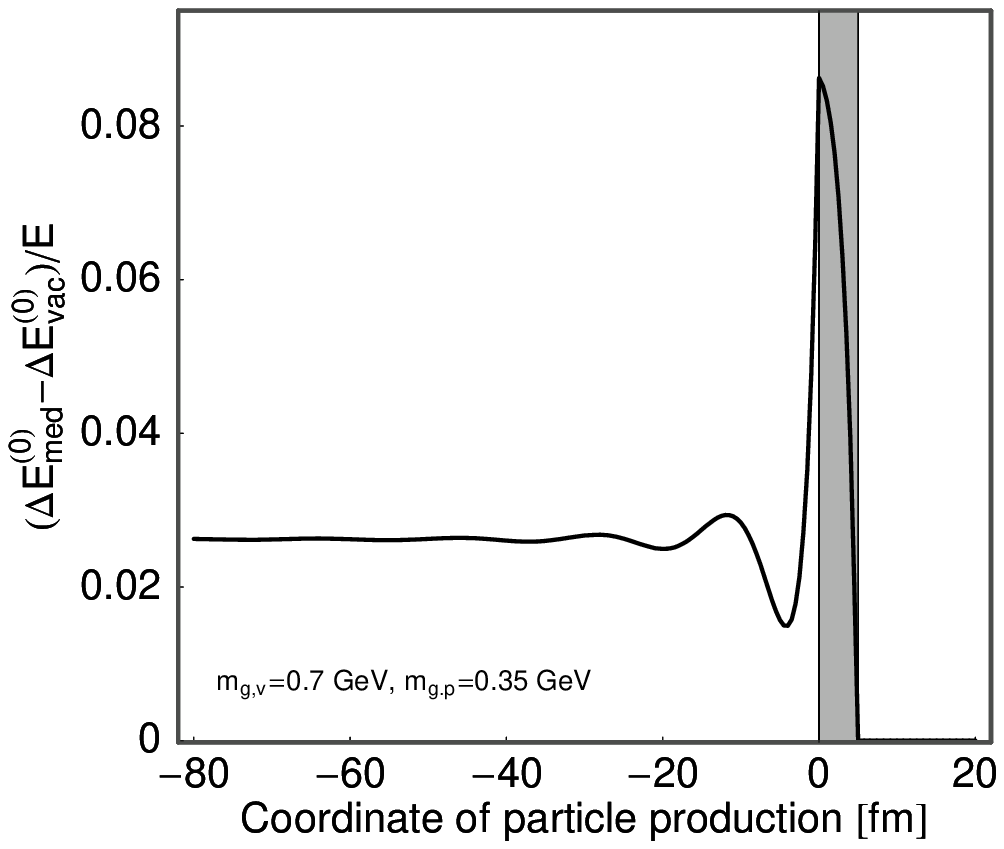}
\caption{ The difference between medium and the vacuum $0^{th}$ order energy 
loss for charm quarks is shown as a function of coordinate of particle 
production. The curves are computed by assuming running coupling given by 
Eq.~(\ref{frozen_alpha}), with different couplings in the medium and 
in the vacuum.  The thickness of the medium is $L=5$~fm. The charm mass is 
$M=1.5$~GeV and the gluon mass in the medium is $m_{g,p} =0.35$~GeV. 
Left (right) panel corresponds to the $m_{g,v}=\Lambda_{QCD}\approx 0.2$ 
($m_{g,v}=0.7$)~GeV case ($m_{g,v}$ is the gluon mass in the vacuum).}
\label{fig:TR4_mgv}
\end{figure}
\vskip 4truemm

To test how the difference between medium and vacuum $0^{th}$ order energy 
loss depends on the different choices of $m_{g,v}$ we first show alternatives 
of Figs.~\ref{fig:TR_LCB},~\ref{fig:TR_Ldep} and~\ref{fig:TR4} for the case 
of $m_{g,v} \approx \Lambda_{QCD}$ and $m_{g,v}\approx 0.7$~GeV (see 
Figs.~\ref{fig:TR_LCB_mgv}, \ref{fig:TR_Ldep_mgv} and~\ref{fig:TR4_mgv} 
respectively). The left panels of 
Figs.~\ref{fig:TR_LCB_mgv},~\ref{fig:TR_Ldep_mgv} and~\ref{fig:TR4_mgv} 
correspond to $m_{g,v} \approx \Lambda_{QCD}$ case. We see that these figures 
are qualitatively similar to the $m_{g,v}=0$ case, although the net effect is 
smaller for $m_{g,v}\approx \Lambda_{QCD}$. This result is expected, since 
$I_p^{(0)}- I_v^{(0)} \propto (m_{g,p}^2-m_{g,v}^2)$, and 
$\Lambda_{QCD}>0$~GeV. On the other hand, the results shown in the right 
panels of Figs.~\ref{fig:TR_LCB_mgv}, \ref{fig:TR_Ldep_mgv}
and~\ref{fig:TR4_mgv} are qualitatively different from the $m_{g,v}=0$ case. 
This is due to the fact that, in this case, the gluon mass in the vacuum is 
larger than in the medium ($m_{g,v}=0.7$~GeV), and thus we would expect that 
$I_p^{(0)}- I_v^{(0)}$ would always be positive definite, in agreement with 
these figures.

\begin{figure}
\vspace*{7.3cm} \includegraphics{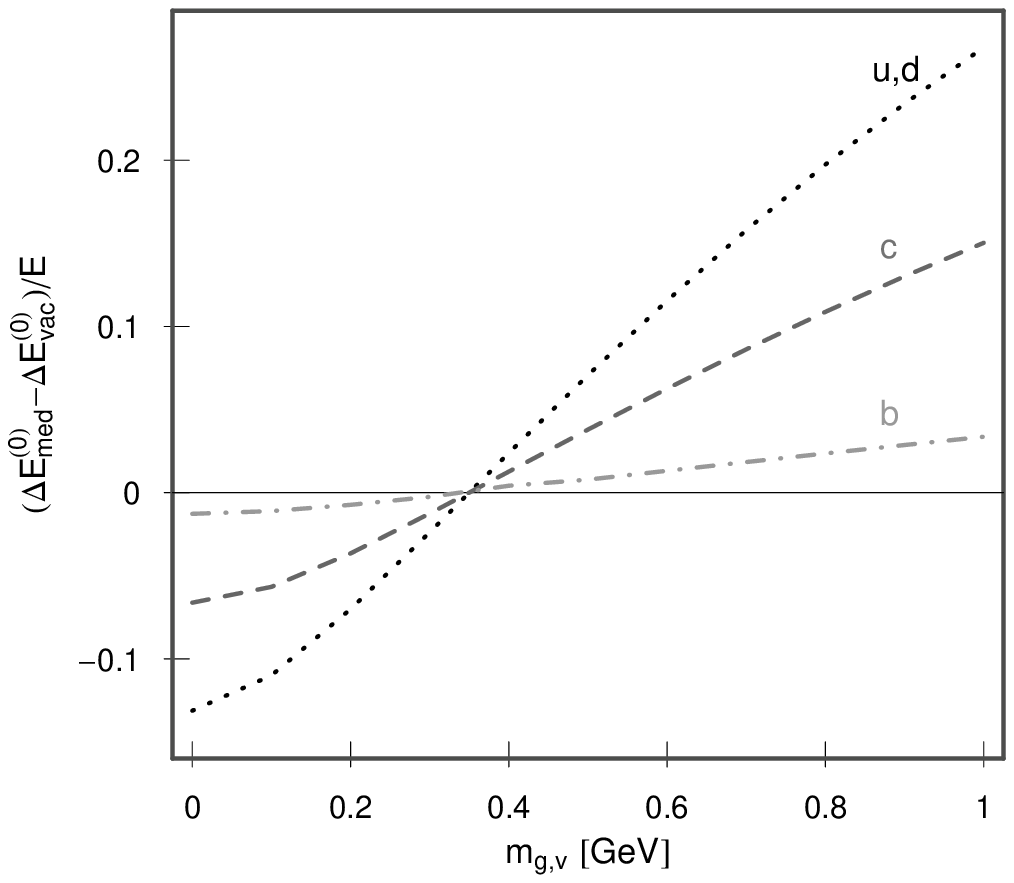}
\caption{ The difference between medium and the vacuum $0^{th}$ order energy 
loss is shown as a function of the gluon mass in the vacuum ($m_{g,v}$). The 
curves are computed by assuming running coupling given by 
Eq.~(\ref{frozen_alpha}), with different couplings in the medium and in the 
vacuum. The dotted curve corresponds to the light quarks, dashed to the charm 
and dot-dashed curve to the bottom quark. The initial energy of the jet is 
$15$~GeV, and thickness of the medium is $L=5$~fm. We take that light, 
charm  and bottom quark masses are $M=0$, $M=1.5$ and $M=4.5$~GeV respectively.}
\label{fig:TR7}
\end{figure}
\vskip 4truemm 

In Fig.~\ref{fig:TR7} we show the difference between medium and the vacuum 
$0^{th}$ order energy loss for light, charm and bottom quarks as a function of 
the vacuum gluon mass ($m_{g,v}$). From this figure we see that for bottom 
quarks the difference is negligible. For charm quarks, in the range of 
experimental interest ($0$~GeV $<m_{g,v}<0.7$~GeV) this difference is in the 
range of about $\pm 5\%$. In~\cite{DGW_proc,DGW_PRL} we showed that this 
difference has small effects on the heavy flavor experimental observables 
(i.e. negligible effect on bottom, and less than $\pm 0.1$ on charm $R_{AA}$). 
However, for the light quarks, we see that the difference between medium and 
the vacuum energy loss is in the range of about $\pm 15\%$, which may have a 
sizable effect on the pion suppression. We therefore conclude that, in
order to obtain consistent predictions for pion suppression data, 1) the gluon 
mass in the vacuum has to be more accurately estimated and 2) the transition 
radiation has to be taken into account.


\section{Net radiative energy loss dependence on transition radiation}
\label{TR5}

In this section we will use the medium induced radiative energy loss 
given in~\cite{Djordjevic:2003zk} to study how the difference between net 
radiative medium ($\Delta E^{(1)}+\Delta E^{(0)}_{med}$) and the vacuum 
($\Delta E^{(0)}_{vac}$) energy loss changes when the transition radiation 
effects are included.   

\subsection{Comparison between light and heavy quark medium induced radiative 
energy loss}

\begin{figure}[h]
\vspace*{6.cm} \includegraphics{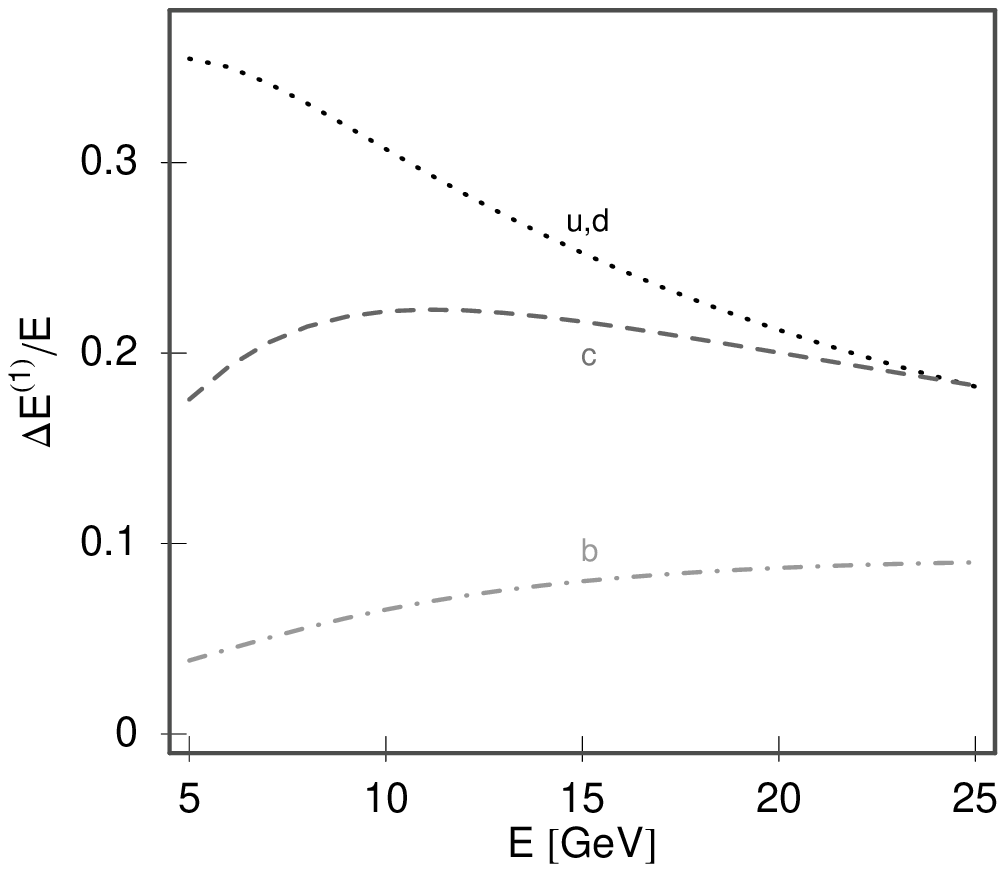}
\includegraphics{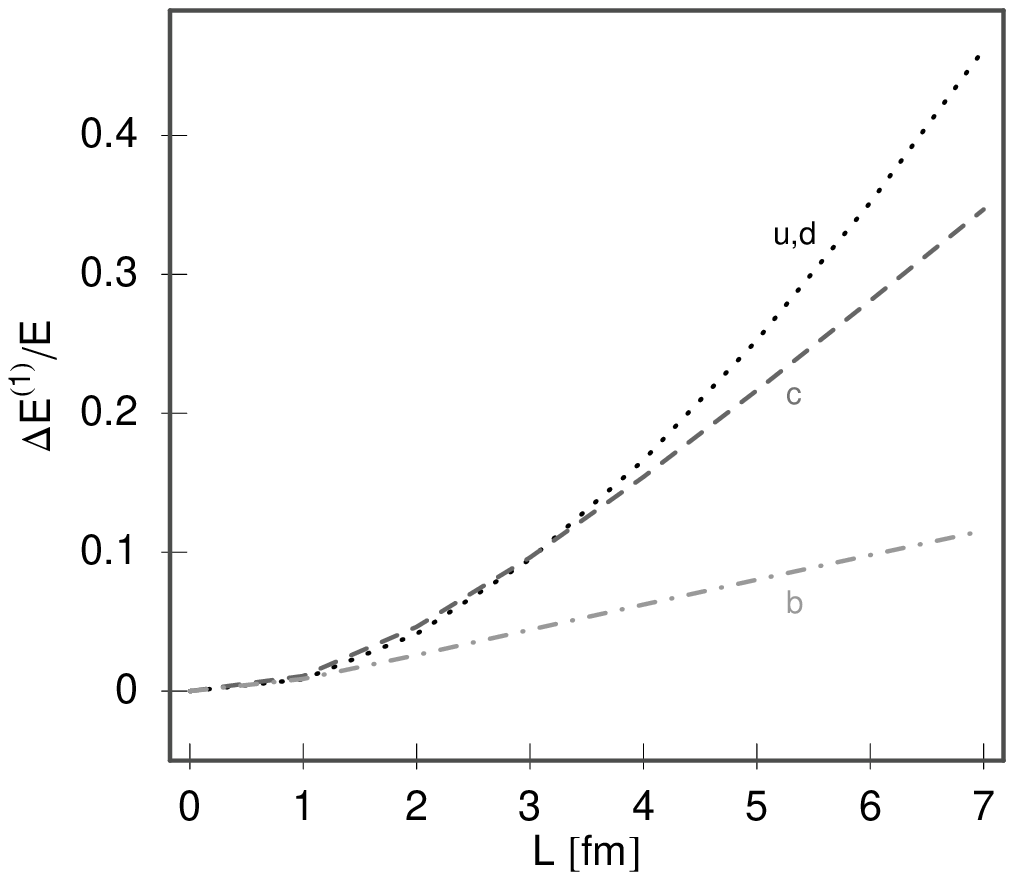}
\caption{ On the left panel the $1^{st}$ order in opacity fractional 
energy loss as a function of initial jet energy is shown for heavy and light 
quark jets. Assumed thickness of the medium is $L=5$~fm. On the right panel 
the $1^{st}$ order in opacity fractional energy loss for a 15~GeV jet
is plotted versus the effective static thickness $L$. Plasma is 
characterized by $m_{g,p}=0.35$~GeV and $\lambda=1$~fm. Dotted curves 
corresponds to light quarks while (dashed) dot-dashed curves
corresponds to charm (bottom).}
\label{FO_reference}
\end{figure}

For the purpose of further comparison, in this subsection, we show the 
$1^{st}$ order medium induced radiative energy loss for light and heavy quarks.
The left panel of Fig.~\ref{FO_reference} shows fractional energy loss to 
$1^{st}$ order in opacity($\Delta E^{(1)}$) as a function of initial jet 
energy. As in the previous sections, we assume running coupling given by 
Eq.~(\ref{frozen_alpha}) with $m_{g}=m_{g,p}=0.35$~GeV. We take $L=5$~fm 
and $\lambda=1$~fm for the plasma parameters. We see that for $5$~GeV jet, the 
finite mass effect leads to a $50 \%$ ($90 \%$) reduction of the energy loss 
for charm (bottom) quarks. On the other hand, for a $20$~GeV jet, we see that 
the finite mass effect has almost no effect on charm quarks while it reduces 
the bottom quark energy loss by $50\%$.

On the right panel of Fig.~\ref{FO_reference} we fix the jet energy to 
$15$~GeV, and look at the fractional energy loss as a function of thickness of 
the medium. Wee see that charm and light quark energy loss dependence is 
similar, while bottom quark remains significantly different and close to the 
linear $L^1$ Bethe-Heitler form. This behavior is expected having in mind the 
left panel in Fig.~\ref{FO_reference}. There we see that, for a $15$~GeV jet, 
the finite mass effect does not have a large influence on charm quarks, while 
it still has a significant influence on heavy bottom quarks.

\subsection{The net radiative energy loss for light and heavy quarks in QCD
  medium}

\begin{figure}[h]
\vspace*{5cm} \includegraphics{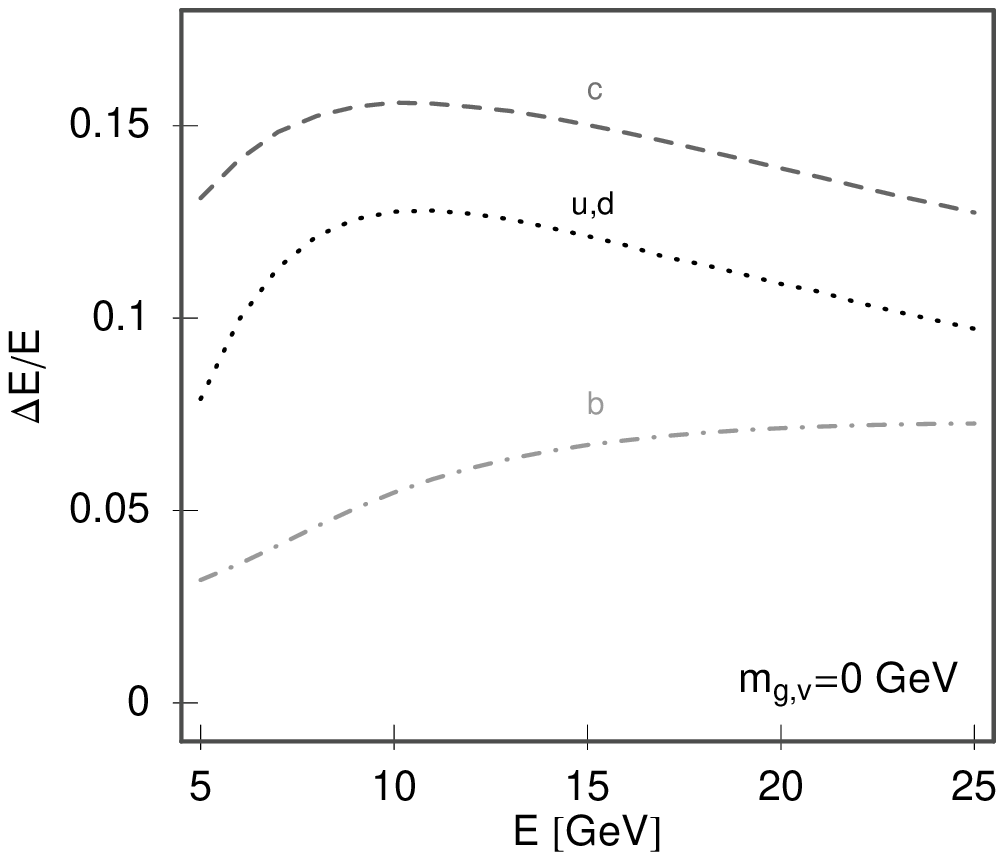}
\includegraphics{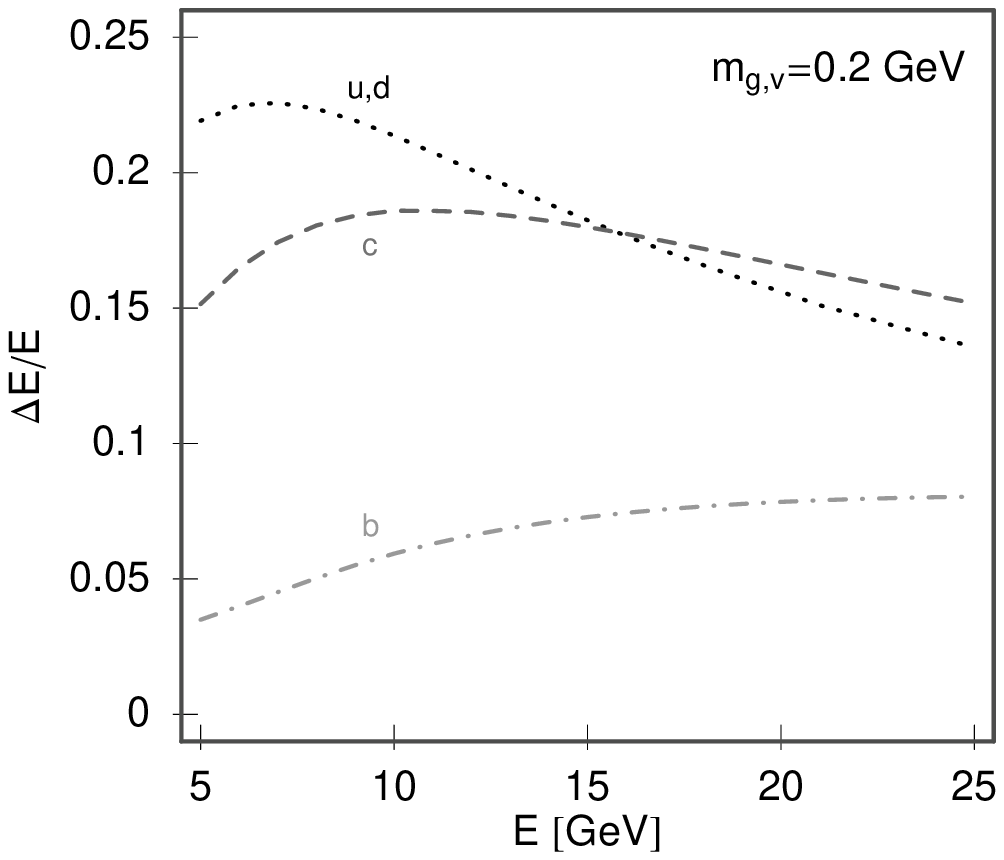}
\includegraphics{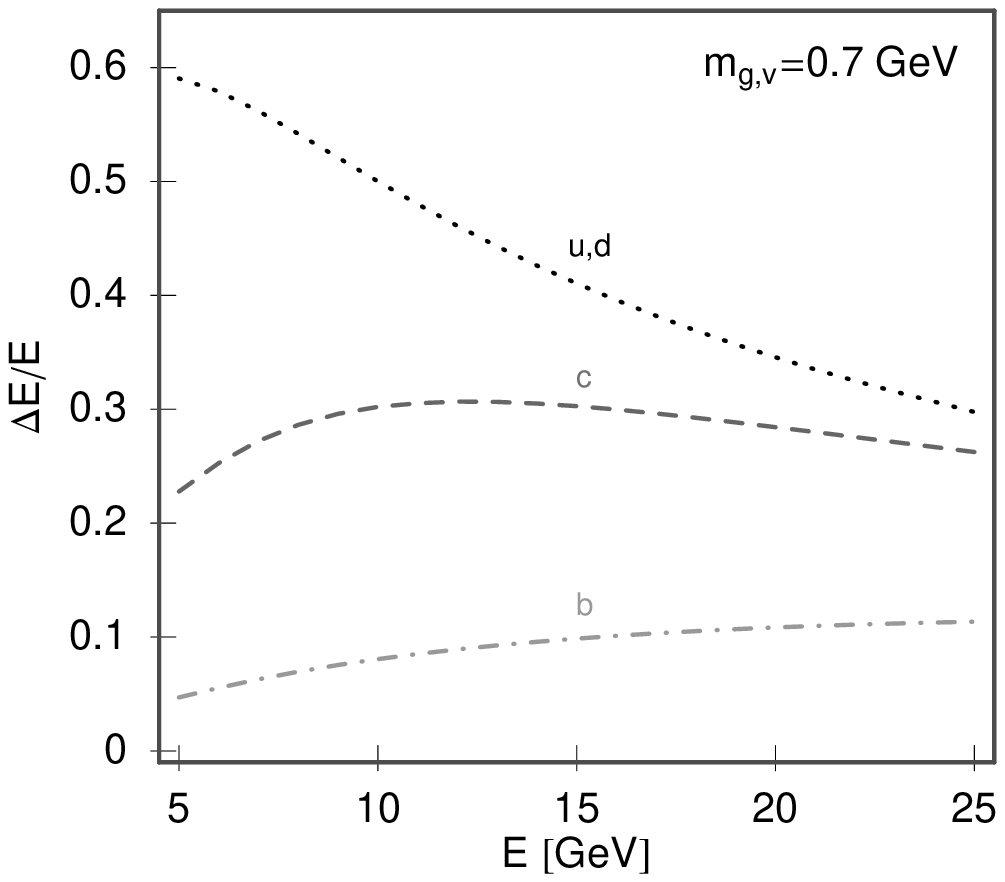}
\caption{ The net radiative fractional energy loss 
($\Delta E=\Delta E^{(1)}+\Delta E^{(0)}_{med}-\Delta E^{(0)}_{vac}$) as a 
function of initial jet energy is shown for heavy and light quark jets. 
Thickness of the medium is $L=5$~fm. Left, central and right panels correspond 
to $m_{g,v}=0, \, 0.2$ and $0.7$~GeV cases respectively. Dotted curves 
corresponds to light quarks, while (dashed) dot-dashed curves corresponds to 
charm (bottom). }
\label{tot_E_dep} 
\end{figure}
\vskip 4truemm 

In this subsection, we first concentrate how the net radiative energy loss 
depends on the initial jet energy for $m_{g,v}=0, \, 0.2$ and $0.7$~GeV 
cases. Figure~\ref{tot_E_dep} was obtained by combining Figs.~\ref{fig:TR_LCB} 
and~\ref{fig:TR_LCB_mgv} with the left panel of Fig.~\ref{FO_reference}. We 
use it to compare the net radiative energy loss results for light, charm and 
bottom quarks. We see that, depending on the gluon mass in the vacuum, the 
transition radiation may either further enhance (for $m_{g,v}>m_{g,p}$) or 
kill the ``dead-cone'' effect (for $m_{g,v}<m_{g,p}$). Additionally, we see 
that the transition radiation may have a significant influence on the net 
radiative light parton energy loss. For example, for the $m_{g,v}=0$~GeV case, the light quark energy loss is smaller than the charm quark energy loss. 
Additionally, in this case, the energy loss for all three types of quarks shows
a weak dependence on the initial jet energy. On the other hand, in the 
$m_{g,v}=0.7$~GeV case, the light quark energy loss is a steeply decreasing 
function of initial jet energy. Additionally, this energy loss is significantly 
larger than both the charm and bottom quark energy losses.  

\begin{figure}[h]
\vspace*{5cm} \includegraphics{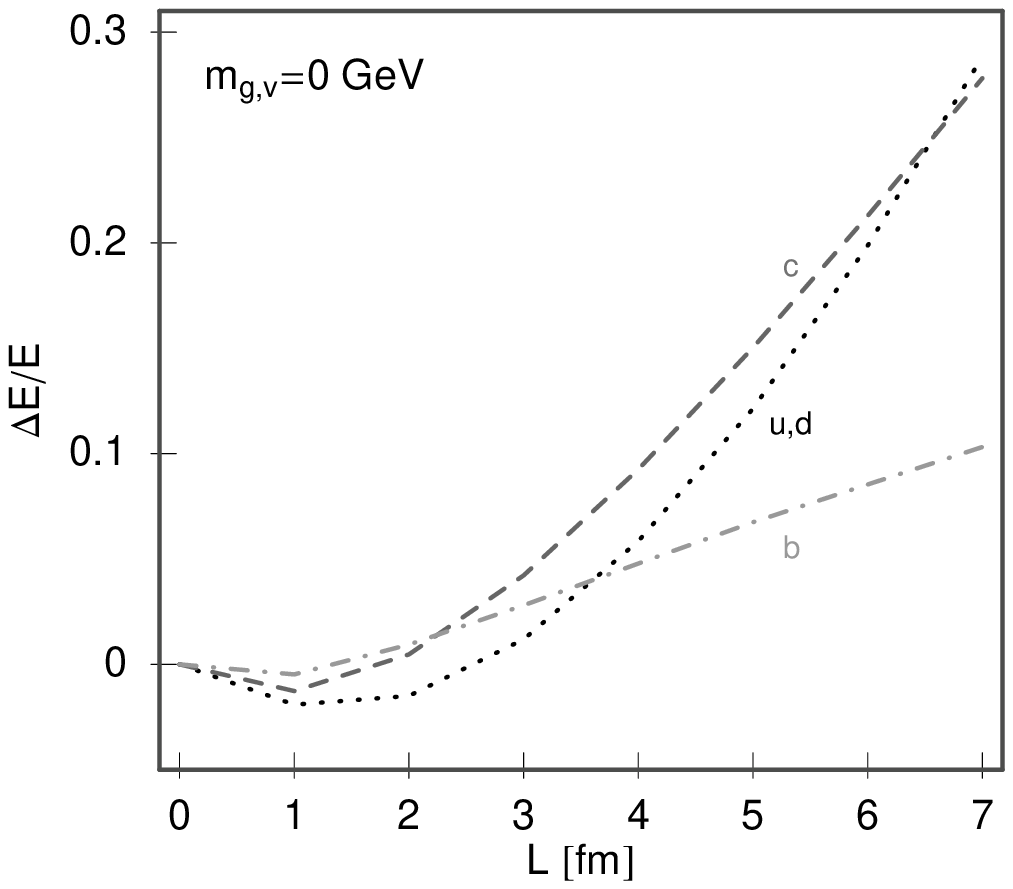}
\includegraphics{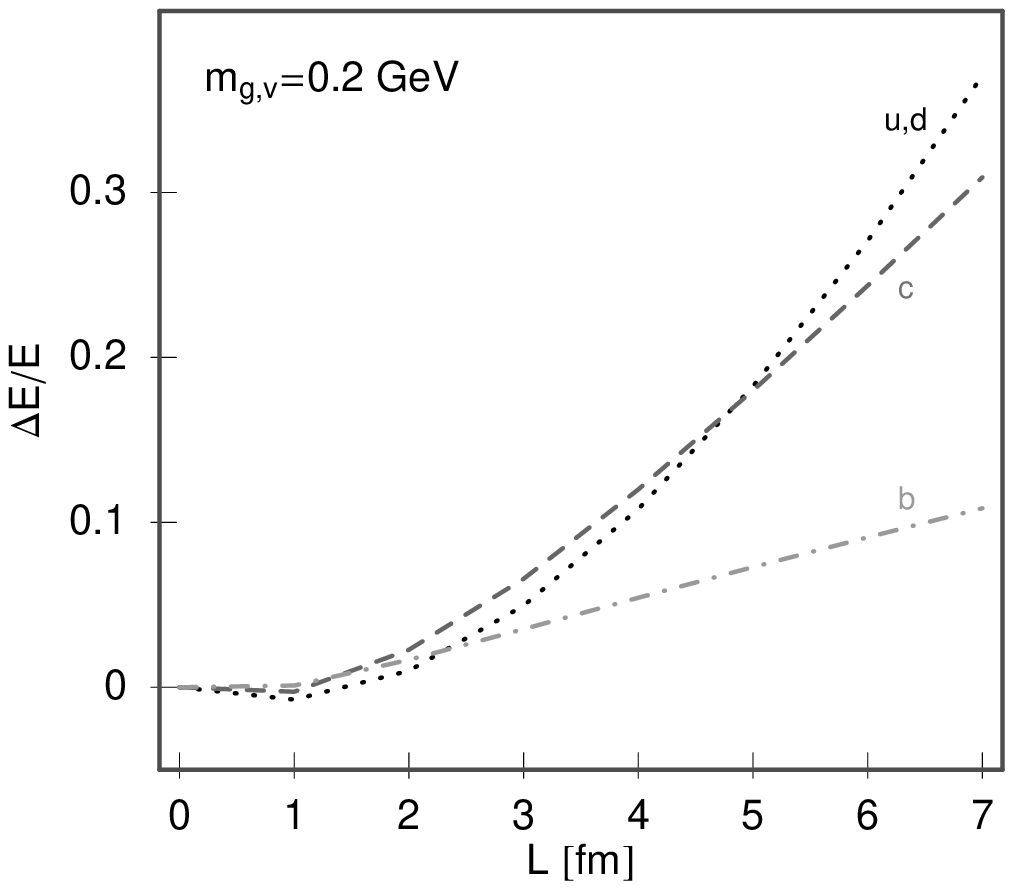}
\includegraphics{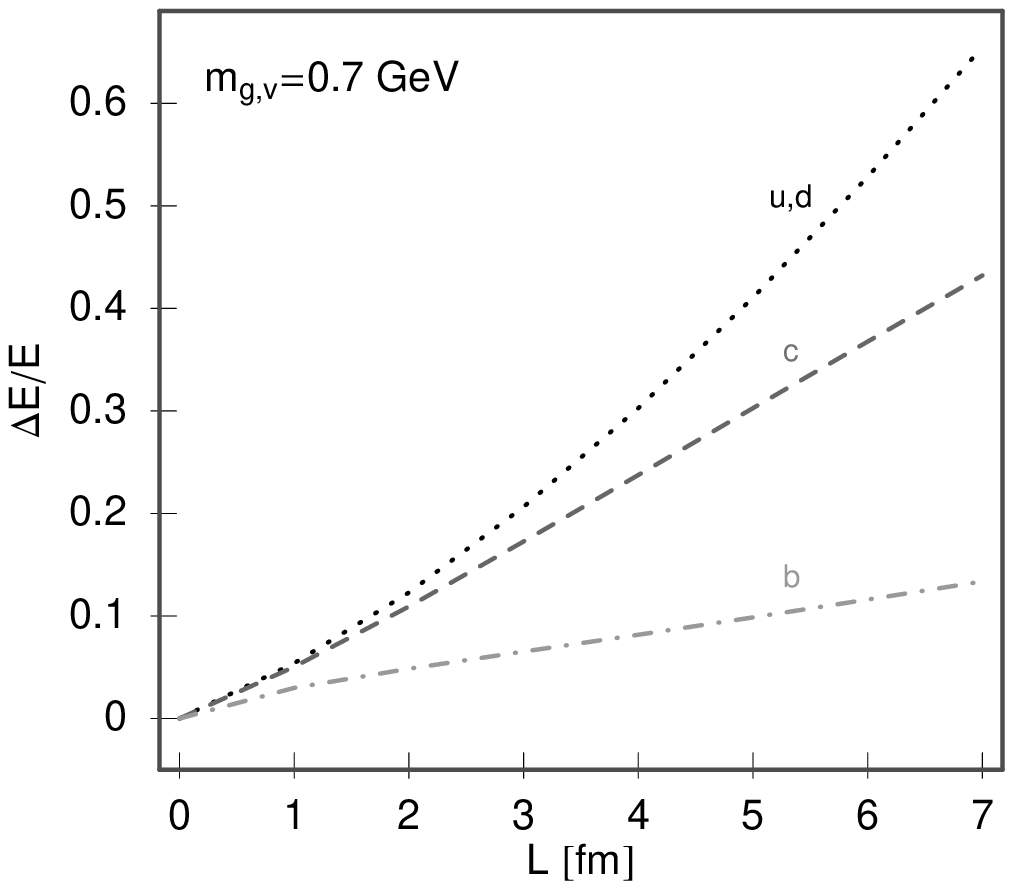}
\caption{ The net radiative fractional energy loss 
($\Delta E=\Delta E^{(1)}+\Delta E^{(0)}_{med}-\Delta E^{(0)}_{vac}$) as a 
function of the thickness of the medium is shown for heavy and light quark 
jets. Initial jet energy is 15~GeV. Left, central and right panels 
correspond to $m_{g,v}=0, \, 0.2$ and $0.7$~GeV cases respectively. Dotted 
curves corresponds to light quarks, while (dashed) dot-dashed curves 
corresponds to charm (bottom). }
\label{tot_L_dep} 
\end{figure}
\vskip 4truemm

Figure~\ref{tot_L_dep} is obtained by combining Figs.~\ref{fig:TR_Ldep} 
and~\ref{fig:TR_Ldep_mgv} with the right panel of Fig.~\ref{FO_reference}. We 
use it to compare the thickness dependence of the net radiative energy loss 
results for light, charm and bottom quarks. In the 
$m_{g,v} \lesssim \Lambda_{QCD}$ case, we see that for the light quarks, 
introduction of the transition radiation may lead to the cancellation of the 
medium induced radiative energy loss for $L \lesssim 3$~fm. This result infers 
that light partons may leave the medium practically unquenched if traveling 
the distances smaller than $3$~fm. Similar energy loss thickness dependence 
was already observed in~\cite{Mioduszewski}, and the Fig.~\ref{tot_E_dep} may 
point how to solve the thickness dependence puzzle posed by~\cite{Mioduszewski}.

\begin{figure}[h]
\vspace*{6cm} \includegraphics{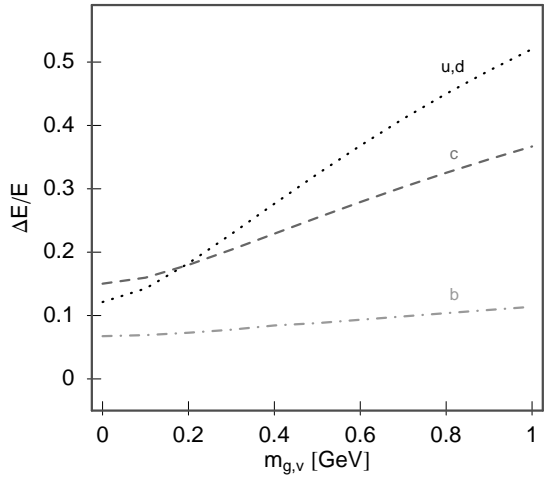}
\caption{ The net radiative fractional energy loss 
($\Delta E=\Delta E^{(1)}+\Delta E^{(0)}_{med}-\Delta E^{(0)}_{vac}$) as a 
function of the gluon mass in the vacuum ($m_{g,v}$) is shown for heavy and 
light quark jets. Dotted curve corresponds to light quarks, 
while (dashed) dot-dashed curve corresponds to charm (bottom). Initial
jet energy is $E=15$~GeV, and thickness of the medium is $L=5$~fm.}\label{tot_mgv_dep} 
\end{figure}
\vskip 4truemm 

Finally in Fig.~\ref{tot_mgv_dep} we fix the jet energy and thickness of the 
QCD medium, and compare the $m_{g,v}$ dependence of the net radiative energy 
loss results for light, charm and bottom quarks. While bottom quark net 
radiative energy loss is independent on $m_{g,v}$, light quark shows 
significant dependence on the on $m_{g,v}$, as expected from 
Fig.~\ref{fig:TR7}. For example, we see that, for the $m_{g,v}<\Lambda_{QCD}$ 
the light quark energy loss is smaller than charm's. On the other hand, in 
$m_{g,v} \approx 1$~GeV case, the difference between light and heavy quark 
energy loss is enhanced from $10 \%$ (see 15~GeV point on 
Fig.~\ref{tot_E_dep}) to approximately $40 \%$.

Unfortunately, at the moment we do not know what value of $m_{g,v}$ would, 
most accurately, reflect confinement effects in the vacuum. However, based on 
Figs.~\ref{tot_E_dep}-\ref{tot_mgv_dep} we see that the $m_{g,v}>m_{g,p}$ 
enhances the differences between the light and heavy quark energy loss results,
and therefore would lead to a significant difference between the suppressions 
of light and heavy parton observables. On the other hand, $m_{g,v}<m_{g,p}$ 
lowers the differences between the light and heavy quark energy losses, and 
would correspondingly lower the differences between the suppressions of light 
and heavy observables. Based on the most recent experimental 
results~\cite{RHIC_single_e}, which suggest similar suppression results for 
pions (light partons observable) and single electrons (heavy quark observable),
we expect that $m_{g,v} \lesssim \Lambda_{QCD}$ is the most appropriate value 
to approximate confinement in the vacuum.   

\section{Conclusions}
\label{TR6}

A finite size medium, with dimensions on the order of the diameter of the 
collided heavy ion, is created in URHIC. Due to that, jets experience a 
transition from medium to the vacuum, which results in additional energy loss,
called transition radiation. Since the Ter-Mikayelian was computed under the 
assumption of an infinite medium~\cite{DG_TM}, in this paper we addressed the 
finite size correction to this effect.

In~\cite{DG_TM} we obtained that, due to the Ter-Mikayelian effect, the medium 
energy loss for charm quarks is reduced by 30 \% compared to the vacuum case. 
We have showed that the finite size correction reduces this effect from 30\% 
to 15\%. The obtained result is intuitively unexpected, since the well known 
QED transition radiation calculations~\cite{TM2,Jackson} give a positive 
difference between the medium and the vacuum energy loss. The discrepancy 
between QED and QCD case results from the fact that in QED experiments a 
particle is produced far outside the medium, and has to cross two vacuum/medium
boundaries in order to reach a detector, while in QCD experiment, the particle 
is produced inside the medium and crosses only one boundary. The QCD effect is 
therefore smaller by (approximately) the energy loss corresponding to one 
boundary crossing.  

Previously, there was a contradiction caused by the fact that the energy loss 
is infinite in the vacuum and finite in the medium, leading to the infinite 
discontinuity between medium and the vacuum energy loss. This problem was long 
avoided by assuming the same zero mass for light partons in both medium and
the vacuum. We here showed that this infinite discontinuity is naturally 
regulated by including transition radiation. To our best knowledge, the work 
presented here is the first consistent solution to this problem.

Further, we showed, that for $m_{g,v} \lesssim \Lambda_{QCD}$, the light parton
may not loose energy when traveling distances smaller than 3~fm. This result 
is similar to the one experimentally observed in~\cite{Mioduszewski}. 
Consequently, one of the future goals is to understand the impact of the 
transition radiation to the light observable's suppression results.
 
We note that our computations were done under the assumption of static medium 
of finite size $L$. Therefore, one of the interesting future problems is 
to study how these results are modified under the influence of the dynamically 
expanding QCD medium with continuously changing density at the edge. We expect 
that for a more gradual density change between the medium and vacuum, the 
difference between medium and vacuum $0^{th}$ order radiation is reduced. 

Finally, we also note that our calculations considered 
only the radiative energy loss and did not take elastic energy loss into 
account. Recent computations by~\cite{Peigne} show that the elastic energy 
loss in the QCD medium is negligible, which supports non-inclusion of the 
elastic energy loss into account. However, in~\cite{Mustafa} it was obtained 
that the elastic energy loss is significant. Consistent inclusion of the 
elastic energy loss into our radiative energy loss formalism is the subject of 
our future work. 

\section*{Acknowledgments} I am grateful to Miklos Gyulassy for valuable 
discussions and critical reading of the manuscript. I also thank Ulrich Heinz
and John Harris for critical reading of the manuscript. Discussions with 
Dimitri Kharzeev, Yuri Kovchegov, James Nagle and Xin-Nian Wang are gratefully 
acknowledged. This work is supported by the Director, Office of Science, 
Office of High Energy and Nuclear Physics, Division of Nuclear Physics, 
of the U.S. Department of Energy under Grants No. DE-FG02-93ER40764
and DE-FG02-01ER41190.

\begin{appendix}

\section{Gluon wave function in finite size QCD medium}

In this appendix we will derive the gluon wave function in a finite size QCD
medium. As already stated in Section~\ref{TR2}, in the spinless case the wave 
function of the emitted gluon with momentum $k$ can be written as (see 
Eq.~(\ref{gluon1}))
\beqar
A_\mu(x) = \epsilon_\mu (k)\;  \Phi_g ({ \mbox x})\; c,
\eeqar{gluon1App}
where $\epsilon(k)=[0,2 \frac{{ \mbox {\scriptsize {\boldmath
	  $\epsilon$}}} {\bf \cdot
      k}}{k^+}, { \mbox {\scriptsize{\boldmath $\epsilon$}}}]$ is the 
transverse polarization and $c$ is the color factor of the radiated 
gluon. In finite size QCD medium, gluon mass (which is proportional to
the temperature) becomes position dependent, and $\Phi_g({ \mbox x})$ 
is the wave function that satisfies the Klein-Gordon equation with 
position dependent gluon mass $m_g({ \mbox x^+})$.

In this section we will derive the wave function $\Phi_g(x)$ in the
light cone coordinate system. To compute this wave function, we have
to solve the fallowing equation
\beqar
(\Box + m_g({ \mbox x^+})^2)\Phi_g(x)=0,
\eeqar{KG1}
where in light cone coordinate system $\Box  \equiv 
4 \,\partial_{{ \mbox {\scriptsize x}^{+}}}
\partial_{{ \mbox {\scriptsize x}^{-}}}-\partial_{{ \mbox {\scriptsize
      {\bf x}}}}^2$. Note that we have an extra factor of $4$ which is
the consequence of the coordinate transformations. Additionally, note
that Klein-Gordon equation now becomes first order in ${ \mbox x^+}$ 
and ${ \mbox x^-}$~\cite{Kogut_1970,Bjorken_1970}.

By assuming that the $\Phi_g({ \mbox x})$ has the fallowing form 
$\Phi_g({ \mbox x})=\phi_1({ \mbox x^+})\phi_2({ \mbox x^-}) 
\, e^{i {\bf k \cdot x}}$, it becomes easy to show that 
\beqar
\phi_2({ \mbox x^-})= e^{-i\frac{1}{2} k^{+} { \mbox {\scriptsize
      x}^{-}}}
\eeqar{phi_1}
where $k^{+}$ is a constant.

Klein-Gordon equation then reduces to
\beqar
\frac{d \ln \phi_1({ \mbox x^+})}{d{ \mbox x^+}}=-i 
\frac{{\bf k}^2+ m_g({ \mbox x^+})^2}{2 k^{+}} = -\frac{i}{2} 
k^{-}({ \mbox x^+}),
\eeqar{phi_2}
where $k^{-}({ \mbox x^+}) \equiv \frac{{\bf k}^2+ m_g({ \mbox {\scriptsize
    x}^{+}})^2}{k^{+}}$. We can now easily obtain the solution for 
$\phi_1({ \mbox x^+})$
\beqar
\phi_1({ \mbox x^+})=e^{-\frac{i}{2} 
\int\limits_{0}^{{ \mbox {\scriptsize x}^{+}}} d\xi \, k^{-}(\xi)},
\eeqar{phi_2_sol}
which, together with Eq.~(\ref{phi_1}), leads to the solution of the gluon
wave function in a finite size QCD medium
\beqar
\Phi_g ({ \mbox x})=e^{-i\frac{1}{2} [k^{+} {\scriptsize{ \mbox x}^{-}}+ 
\int\limits_0^{{\scriptsize{ \mbox x^+}}} d\xi \, 
k^{-}(\xi)] + i {\bf k \cdot x}}
\eeqar{Phi_app}

This solution is valid for arbitrary gluon mass functional dependence
$m_g({ \mbox x^+})$, and is not limited to the static medium case which we 
consider in Section~\ref{TR2}. In a static medium, where $m_g({
  \mbox x^+})$ is given by Eq.~(\ref{mg}), the Eq.~(\ref{Phi_app})
reduces to
\beqar
\Phi_g ({ \mbox x})=e^{-ik_p { \mbox{\scriptsize x}}}\; 
\theta(L-\frac{{ \mbox x}^+}{2}) + e^{-i(k^-_p-k^-_v)L} \, 
e^{-ik_v { \mbox{\scriptsize x}}}\; \theta(\frac{{ \mbox x}^+}{2}-L)
\eeqar{Phi_stat}
where 
\beqar
k_p&=&[k^+,\frac{{\bf k}^2+m_{g,p}^2}{k^+},{\bf k}] \; ,\nonumber \\
k_v&=&[k^+,\frac{{\bf k}^2+m_{g,v}^2}{k^+},{\bf k}]
\eeqar{gluon_momentum} 
is the gluon momentum in the medium and the vacuum respectively.

\section{Computation of $M_{rad}$ in the case when the jet is
produced inside the medium}

In this appendix we will compute the amplitude of the diagram $M_{rad}$ in 
the case when the jet is produced inside a static QCD medium of size $L$. 
To do that we start with the Eq.~(\ref{eq:1}), i.e.

\beqar
M^{rad} =  \int d^4{ \mbox x_0} \; J({ \mbox x_0}) \; d^4 { \mbox x_1}
\; \Delta_M ({ \mbox x_1}-{ \mbox x_0}) \; v^{\mu}({ \mbox x_1}) \; 
A_{\mu}^{\dagger}({ \mbox x_1}) \Phi^{\dagger}({ \mbox x_1}).
\eeqar{eq:1app}
Here $\Phi({ \mbox x_1})=e^{-ip{ \mbox {\scriptsize x}}}$ is the wave
function of the final quark with (on-shell) momentum $p$ and 
$A_\mu({ \mbox x_1})$ is the wave function of the emitted gluon. 
Vertex function $v^{\mu}({ \mbox x_1})$ is given by $v^{\mu}({ \mbox x_1})
=g({ \mbox x_1})(\overleftarrow{\partial}^{\mu}-
\overrightarrow{\partial}^{\mu})$, where $g({ \mbox x_1})$ is the
running coupling constant. 

After replacing Eqs.~(\ref{propagator})-(\ref{Phi}) in
Eq.(\ref{eq:1app}), we obtain (note that since x$_1>$ x$_0$, 
we keep only the first term in Eq.~(\ref{propagator}))

\beqar
M_{rad} &=&  \int d^4{ \mbox x_0} \; J({ \mbox x_0}) \; d^4 { \mbox x_1}
\; \frac{-i}{(2 \pi)^3} \int \frac{dp'^+ d^2 {\bf p'}}{2 p'^+} 
\theta(({ \mbox x}_1-{ \mbox x}_0)^+) \;
e^{-ip'( {\mbox {\scriptsize x}}_1-{\mbox {\scriptsize x}}_0)} \nonumber \\
&& \hspace*{0.5cm} g({ \mbox x}_1)
(\overleftarrow{\partial}^{\mu}_{{ \mbox {\scriptsize x}}_1}-
\overrightarrow{\partial}^{\mu}_{{\mbox {\scriptsize x}}_1})\;
\epsilon_\mu (k)\,  \Phi^*_g ({ \mbox x_1}) \, c \; 
e^{ip{ \mbox {\scriptsize x}}_1}  \nonumber \\
&=& \int \frac{dp'^+ d^2 {\bf p'}}{2 p'^+} \; \int d^4{ \mbox x_0} \; 
J({ \mbox x_0}) e^{i (p+k) { \mbox {\scriptsize x}}_0} \;  
\frac{-i}{(2\pi)^3} \; \int d^4 { \mbox x_1} (-i) g({ \mbox x}_1)
\nonumber \\ && \hspace*{0.5cm} 
(p'+p)^\mu \epsilon_\mu (k)\, e^{i(p-p')({ \mbox {\scriptsize x}}_1-
{\mbox {\scriptsize x}}_0)} 
\Phi^*_g ({ \mbox x}_1-{ \mbox x}_0) \theta(({ \mbox x}_1-{ \mbox
  x}_0)^+) \; c.
\eeqar{eq:2app}

In the static medium we can replace $\Phi_g({ \mbox x})$ by 
Eq.~(\ref{Phi_stat}), and the Eq.~(\ref{eq:2app}) reduces to (note 
${\mbox x} ={ \mbox x}_1-{\mbox x}_0$)

\beqar
M_{rad} &=& \int \frac{dp'^+ d^2 {\bf p'}}{2 p'^+} \; J(p+k) \;  
(2 p' \cdot \epsilon)\nonumber \\  
&& \hspace*{0.5cm}
\{ \frac{-i}{(2\pi)^3} \;  \int d^4 { \mbox x}\; (-i) g_p\; 
e^{i(p+k_p-p'){ \mbox {\scriptsize  x}}}\; \theta({ \mbox x}^+)
\theta(L-\frac{{ \mbox x}^+}{2}) \nonumber \\ && \hspace*{0.5cm} \;+
\; e^{i(k^-_p-k^-_v)L} \,\frac{-i}{(2\pi)^3} \;  \int d^4 { \mbox x}\; 
(-i) g_v\; e^{i(p+k_v-p'){ \mbox {\scriptsize  x}}}\; 
\theta(\frac{{ \mbox x}^+}{2}-L) \} \; c \;  ,
\eeqar{eq:3app}
where we used $J(p+k)=\int d^4{ \mbox x_0} \; J({ \mbox x_0}) 
e^{i (p+k) { \mbox {\scriptsize x}}_0}$.

We will first compute $I_1=\frac{-i}{(2\pi)^3} \;  \int d^4 { \mbox x}\; 
(-i) g_p\; e^{i(p+k_p-p'){ \mbox {\scriptsize  x}}}\; \theta({ \mbox x}^+)
\theta(L-\frac{{ \mbox x}^+}{2})$. Note that in the light cone gauge
$d^4 { \mbox x}= 1/2 \, d{ \mbox x}^+  d{ \mbox x}^- d{\bf x}$,
leading to
\beqar
I_1&=&\frac{-g_p}{(2\pi)^3}\; \int 1/2 \, d{ \mbox x}^+  d{ \mbox x}^-
d{\bf x}\; e^{\frac{i}{2}(p+k-p')^+{ \mbox {\scriptsize  x}}^-}
\; e^{\frac{i}{2}(p+k_p-p')^-{ \mbox {\scriptsize  x}}^+}  
e^{-i {\bf (p+k-p') \cdot x}}\; \theta({ \mbox x}^+)
\theta(L-\frac{{ \mbox x}^+}{2})\nonumber \\ 
&=&-g_p\; \delta (p^+ + k^+ -p'^+) \delta({\bf p+k-p'})
\int_0^{2L} d{ \mbox x}^+\; 
e^{\frac{i}{2}(p+k_p-p')^-{ \mbox {\scriptsize  x}}^+}\nonumber \\ 
&=&-2ig_p\; \delta (p^+ + k^+ -p'^+) \delta({\bf p+k-p'})
\frac{1-e^{i(p+k_p-p')^- L}}{(p+k_p-p')^-} 
\eeqar{eq_I1}
Without loss of generality, we can take for the initial quark the 
plane wave state in the $ {\bf x}$-plane and set ${\bf p'}=0$. Then 
${\bf p}=- {\bf k}$, and $(p+k_p-p')^-=\chi_p$, where $\chi_p$ is given by 
Eq.~(\ref{eq:chi}). By using this, the Eq.~(\ref{eq_I1}) finally
reduces to
\beqar
I_1=-2ig_p\; \delta (p^+ + k^+ -p'^+) \delta({\bf p+k-p'})
\frac{1-e^{i\chi_p L}}{\chi_p} 
\eeqar{eq_I1_f}

In the same way $I_2= e^{i(k^-_p-k^-_v)L} \,\frac{-i}{(2\pi)^3} \;  
\int d^4 { \mbox x}\; (-i) g_v\; e^{i(p+k_v-p'){ \mbox {\scriptsize  x}}}\; 
\theta(\frac{{ \mbox x}^+}{2}-L)$ reduces to
\beqar
I_2&=&-2ig_p\; e^{-i(k^-_p-k^-_v)L} \delta (p^+ + k^+ -p'^+) 
\delta({\bf p+k-p'}) \frac{e^{i \chi_v L}}{\chi_v}\nonumber \\
&=&-2ig_p\; \delta (p^+ + k^+ -p'^+) 
\delta({\bf p+k-p'}) \frac{e^{i \chi_p L}}{\chi_v} 
\eeqar{eq_I2_f}

After we plug in the Eqs.~(\ref{eq_I1_f}) and~(\ref{eq_I2_f}) into 
Eq.~(\ref{eq:2app}) and use $(p' \cdot \epsilon) = 
\frac{{\mbox {\scriptsize {\boldmath $\epsilon$}}}{\bf \cdot k}}{x}$ 
($x \equiv \frac{k^+}{p'^+}$), the $M_{rad}$ reduces to

\beqar
M_{rad} &=& -2 i \int {dp'^+ d^2 {\bf p'}} \; J(p+k) \;  
(p' \cdot \epsilon) \;\delta (p^+ + k^+ -p'^+)\; \delta({\bf p+k-p'}) 
\nonumber \\  &&\; ( \frac{g_p}{p'^+}\frac{1-e^{i \chi_p L}}{\chi_p} +
\frac{g_v}{p'^+}\frac{e^{i \chi_p L}}{\chi_v}) \; c \nonumber \\
&=&-2 i\, J(p+k)\frac{{\mbox{\boldmath $\epsilon$}}{\bf \cdot k}}{x} 
\left [\frac{g_p}{p'^+}\frac{1-e^{i\chi_p L}}{\chi_p} +
\frac{g_v}{p'^+}\frac{e^{i\chi_p L}}{\chi_v} \right] c \; ,
\eeqar{M_rad_f}
which is the Eq.~(\ref{MradStat}) given in section~\ref{TR2}.

\section{Computation of $M_{rad}$ in the case when the jet is produced 
outside the medium}

In this appendix we will compute the Eq.~(\ref{eq:2app}) in the
case when the jet is produced at the distance $l_0$ from the medium, i.e. 
at $\mbox{x}_0^+=-2l_0<0$. Then, for $\mbox{x}=\mbox{x}_1-\mbox{x}_0$,
(where $\mbox{x}_1$ is the gluon production point), $m_g({ \mbox x^+})$ 
can be written as
\beq
m_g({ \mbox x^+})=m_{g,v} \; \theta({\mbox x^+})
\theta(l_0-\frac{{\mbox x^+}}{2})+
m_{g,p} \; \theta(\frac{{\mbox x^+}}{2}-l_0)
\theta(l_0+L-\frac{{ \mbox x^+}}{2})+ 
m_{g,v} \; \theta(\frac{{ \mbox x^+}}{2}-(L+l_0))
\eeq{mg_l0}

By using Eq.~(\ref{mg_l0}), $\Phi_g ({ \mbox x})$ (see Eq.~(\ref{Phi_app})) 
reduces to
\beqar
\Phi_g ({ \mbox x})&=&e^{-ik_v { \mbox{\scriptsize x}}}\;
\theta({\mbox x^+})\theta(l_0-\frac{{\mbox x^+}}{2}) + 
e^{-i(k^-_v-k^-_p)l_0}e^{-ik_p { \mbox{\scriptsize x}}}\; 
\theta(\frac{{\mbox x^+}}{2}-l_0)\theta(l_0+L-\frac{{ \mbox x^+}}{2})
\nonumber \\
&& \hspace*{3.9cm}+e^{-i(k^-_p-k^-_v)L} e^{-ik_v { \mbox{\scriptsize x}}}\; 
\theta(\frac{{ \mbox x^+}}{2}-(L+l_0))
\eeqar{Phi_stat_l0}

With the use of Eq.~(\ref{Phi_stat_l0}), the Eq.~(\ref{eq:2app}) reduces to 

\beqar
M_{rad} &=& \int \frac{dp'^+ d^2 {\bf p'}}{2 p'^+} \; J(p+k) \;  
(2 p' \cdot \epsilon)\nonumber \\  
&& \hspace*{0.5cm}
\{ \frac{-i}{(2\pi)^3} \;  \int d^4 { \mbox x}\; (-i) g_v\; 
e^{i(p+k_v-p'){ \mbox {\scriptsize  x}}}\; \theta({ \mbox x}^+)
\theta(l_0-\frac{{ \mbox x}^+}{2}) \nonumber \\ && \hspace*{0.5cm} \;+
\; e^{i(k^-_v-k^-_p)l_0} \,\frac{-i}{(2\pi)^3} \;  \int d^4 { \mbox x}\; 
(-i) g_p\; e^{i(p+k_p-p'){ \mbox {\scriptsize  x}}}\; 
\theta(\frac{{\mbox x^+}}{2}-l_0)\theta(l_0+L-\frac{{ \mbox x^+}}{2})
\nonumber \\ && \hspace*{0.5cm} \;+
\; e^{i(k^-_p-k^-_v)L} \,\frac{-i}{(2\pi)^3} \;  \int d^4 { \mbox x}\; 
(-i) g_v\; e^{i(p+k_v-p'){ \mbox {\scriptsize  x}}}\; 
\theta(\frac{{ \mbox x^+}}{2}-(L+l_0)) \} \; c \; .
\eeqar{Mrad_l0_1}

By applying the same procedure as in Appendix B, $M_{rad}$ finally
reduces to 

\beqar
M_{rad} &=& -2 i \int {dp'^+ d^2 {\bf p'}} \; J(p+k) \;  
(p' \cdot \epsilon) \;\delta (p^+ + k^+ -p'^+)\; \delta({\bf p+k-p'}) 
\nonumber \\  &&\hspace*{1cm}\; ( \frac{g_v}{p'^+}
\frac{1-e^{i \chi_v l_0}}{\chi_v} +
\frac{g_p}{p'^+} e^{i(k^-_v-k^-_p)l_0} 
\frac{e^{i\chi_p l_0}-e^{i\chi_p (L+l_0)}}{\chi_p}\nonumber \\
&& \hspace*{3.55cm}+ \frac{g_v}{p'^+} \;
e^{i(k^-_p-k^-_v)L}\frac{e^{i\chi_v (L+l_0)}}{\chi_v}) \; c \nonumber \\
&=& -2 i \int {dp'^+ d^2 {\bf p'}} \; J(p+k) \;  
(p' \cdot \epsilon) \;\delta (p^+ + k^+ -p'^+)\; \delta({\bf p+k-p'})
\nonumber \\ && \hspace*{1cm}\; ( \frac{g_v}{p'^+}\frac{1-e^{i \chi_v
    l_0}} {\chi_v} +\frac{g_p}{p'^+} 
\frac{e^{i\chi_v l_0}-e^{i(\chi_p L+\chi_v l_0)}}{\chi_p}
+ \frac{g_v}{p'^+} \;
\frac{e^{i(\chi_p L+\chi_v l_0)}}{\chi_v}) \; c \nonumber \\
&=&-2 i\, J(p+k)\frac{{\mbox{\boldmath $\epsilon$}}{\bf \cdot k}}{x} 
\frac{1}{p'^+}\left [\frac{g_v}{\chi_v} -(\frac{g_v}{\chi_v}-
\frac{g_p}{\chi_p})e^{i\chi_v l_0}(1-e^{i\chi_p L})\right ]
\eeqar{M_rad_l0_f}
which is the Eq.~(\ref{Mrad_outside}) given in section~\ref{TR3}.

\end{appendix}

\end{document}